\documentclass[twocolumn]{aastex631} 

\usepackage{CLASSY_Preamble}

\usepackage{amssymb}	
\usepackage{longtable}
\usepackage{scalerel}

\mathchardef\mhyphen="2D

\newcommand{\oii}{O\,{\sc ii}}
\newcommand{\oiii}{O\,{\sc iii}}

\newcommand{\cii}{[C\,{\sc ii}]}

\newcommand{\mgii}{Mg\,{\sc ii}}

\newcommand{\angstrom}{\text{ \normalfont\AA}}

\mathchardef\mhyphen="2D

\def\lya{Ly$\alpha$}
\def\ly{$\lambda$}

\def\hi{H\,{\sc i}}
\def\hii{H\,{\sc ii}}

\def\cii{C\,{\sc ii}}

\def\oi{O\,{\sc i}}
\def\oiii{O\,{\sc iii}}

\def\mgii{Mg\,{\sc ii}}

\def\Siii{Si\,{\sc ii}}

\def\Q0059{Q0059--2735}

\def\S2S3{S2S3}

\definecolor{blk}{rgb}{0.0,0.0,0.0}
\definecolor{red}{rgb}{0.75,0.0,0.0}
\definecolor{yel}{rgb}{0.65,0.65,0.0}
\definecolor{grn}{rgb}{0.0,0.75,0.0}
\definecolor{blu}{rgb}{0.0,0.0,0.75}
\definecolor{gry}{rgb}{0.75,0.75,0.75}

\def\nh{\ifmmode n_\mathrm{\scriptscriptstyle H} \else $n_\mathrm{\scriptscriptstyle H}$\fi}
\def\ne{\ifmmode n_\mathrm{\scriptstyle e} \else $n_\mathrm{\scriptstyle e}$\fi}
\def\Te{\ifmmode T_\mathrm{\scriptstyle e} \else $T_\mathrm{\scriptstyle e}$\fi}
\def\Qh{\ifmmode Q_\mathrm{\scriptstyle H} \else $Q_\mathrm{\scriptstyle H}$\fi}
\def\Uh{\ifmmode U_\mathrm{\scriptstyle H} \else $U_\mathrm{\scriptstyle H}$\fi}
\def\Nh{\ifmmode N_\mathrm{\scriptstyle H} \else $N_\mathrm{\scriptstyle H}$\fi}
\def\Uhhp{\ifmmode U_\mathrm{\scriptstyle H,HP} \else $U_\mathrm{\scriptstyle H,HP}$\fi}
\def\Nhhp{\ifmmode N_\mathrm{\scriptstyle H,HP} \else $N_\mathrm{\scriptstyle H,HP}$\fi}
\def\Uhvhp{\ifmmode U_\mathrm{\scriptstyle H,VHP} \else $U_\mathrm{\scriptstyle H,VHP}$\fi}
\def\Nhvhp{\ifmmode N_\mathrm{\scriptstyle H,VHP} \else $N_\mathrm{\scriptstyle H,VHP}$\fi}
\def\Nion{\ifmmode N_\mathrm{\scriptstyle ion} \else $N_\mathrm{\scriptstyle ion}$\fi}

\def\Zsun{\ifmmode {\rm Z}_{\odot} \else $Z_{\odot}$\fi}
\def\Msun{\ifmmode {\rm M}_{\odot} \else M$_{\odot}$\fi}
\def\kms{\ifmmode {\rm km~s}^{-1} \else km~s$^{-1}$\fi}
\def\Lya{\ifmmode {\rm Ly}\alpha \else Ly$\alpha$\fi}
\def\Lyb{\ifmmode {\rm Ly}\beta \else Ly$\beta$\fi}
\def\Lyg{\ifmmode {\rm Ly}\gamma \else Ly$\gamma$\fi}
\def\Lyd{\ifmmode {\rm Ly}\delta \else Ly$\delta$\fi}
\def\neaod{\ifmmode n_\mathrm{\scriptscriptstyle AOD} \else $n_\mathrm{\scriptscriptstyle AOD}$\fi}
\def\necrit{\ifmmode n_\mathrm{\scriptstyle cr} \else $n_\mathrm{\scriptstyle cr}$\fi}
\def\ncr{\ifmmode n_\mathrm{\scriptstyle cr} \else $n_\mathrm{\scriptstyle cr}$\fi}
\def\nepi{\ifmmode n_\mathrm{\scriptscriptstyle PI} \else $n_\mathrm{\scriptscriptstyle PI}$\fi}
\def\gtorder{\mathrel{\raise.3ex\hbox{$>$}\mkern-14mu\lower0.6ex\hbox{$\sim$}}}
\def\ltorder{\mathrel{\raise.3ex\hbox{$<$}\mkern-14mu\lower0.6ex\hbox{$\sim$}}}

\def\vro{\ifmmode v_\mathrm{\scriptscriptstyle 1, \scriptstyle r} \else $v_\mathrm{\scriptscriptstyle 1, \scriptstyle r}$\fi}
\def\vrc{\ifmmode v_\mathrm{\scriptscriptstyle 2, \scriptstyle r} \else $v_\mathrm{\scriptscriptstyle 2, \scriptstyle r}$\fi}
\def\vzo{\ifmmode v_\mathrm{\scriptscriptstyle 1, \scriptstyle z} \else $v_\mathrm{\scriptscriptstyle 1, \scriptstyle z}$\fi}
\def\vzc{\ifmmode v_\mathrm{\scriptscriptstyle 2, \scriptstyle z} \else $v_\mathrm{\scriptscriptstyle 2, \scriptstyle z}$\fi}

\newcommand{\fescLyC}{\textit{f}$_{\text{esc}}^{\text{ LyC}}$}
\newcommand{\fescLyA}{\textit{f}$_{\text{esc}}^{\text{ Ly}\alpha}$}
\newcommand{\fescMgII}{\textit{f}$_{\text{esc}}^{\text{ MgII}}$}

\newcommand{\fescLyCPred}{\textit{f}$_{\text{esc,pd}}^{\text{ LyC}}$}

\newcommand{\RMg}{$R$}
\newcommand{\MgE}{MgE}
\newcommand{\nMgE}{non--MgE}

\newcommand{\CF}{C$_{f}$(\mgii)}
\newcommand{\CFHI}{C$_{f}$(\hi)}
\newcommand{\Tthin}{$\tau_{\text{thin}}$}
\newcommand{\Tthick}{$\tau_{\text{thick}}$}
\newcommand{\Athin}{$A_{\text{thin}}$}
\newcommand{\Athick}{$A_{\text{thick}}$}

\def\ZStar{\ifmmode {\rm Z}_\text{stars} \else $Z_\text{stars}$\fi}
\def\ZGas{\ifmmode {\rm Z}_\text{gas} \else $Z_\text{gas}$\fi}

\begin{document}

\submitjournal{AASJournal ApJ}
\shortauthors{Xu et al.}
\shorttitle{\mgii\ Properties in Low-z Lyman Continuum Survey}

\title{The Low-Redshift Lyman Continuum Survey:\\ Optically Thin and Thick \mgii\ Lines as Probes of Lyman Continuum Escape}

\author[0000-0002-9217-7051]{Xinfeng Xu}
\affiliation{Center for Astrophysical Sciences, Department of Physics \& Astronomy, Johns Hopkins University, Baltimore, MD 21218, USA}

\author[0000-0002-6586-4446]{Alaina Henry}
\affiliation{Center for Astrophysical Sciences, Department of Physics \& Astronomy, Johns Hopkins University, Baltimore, MD 21218, USA}
\affiliation{Space Telescope Science Institute, 3700 San Martin Drive, Baltimore, MD 21218, USA}

\author[0000-0001-6670-6370]{Timothy Heckman}
\affiliation{Center for Astrophysical Sciences, Department of Physics \& Astronomy, Johns Hopkins University, Baltimore, MD 21218, USA}

\author[0000-0002-0302-2577]{John Chisholm}
\affiliation{Department of Astronomy, The University of Texas at Austin, 2515 Speedway, Stop C1400, Austin, TX 78712, USA}

\author[0000-0001-8442-1846]{Rui Marques-Chaves}
\affiliation{Department of Astronomy, University of Geneva, 51 Chemin Pegasi, 1290 Versoix, Switzerland}

\author[0000-0002-6085-5073]{Floriane Leclercq}
\affiliation{Department of Astronomy, The University of Texas at Austin, 2515 Speedway, Stop C1400, Austin, TX 78712, USA}

\author[0000-0002-4153-053X]{Danielle A. Berg}
\affiliation{Department of Astronomy, The University of Texas at Austin, 2515 Speedway, Stop C1400, Austin, TX 78712, USA}

\author[0000-0002-6790-5125]{Anne Jaskot}
\affiliation{Department of Astronomy, Williams College, Williamstown, MA 01267, United States}

\author[0000-0001-7144-7182]{Daniel Schaerer}
\affiliation{Department of Astronomy, University of Geneva, 51 Chemin Pegasi, 1290 Versoix, Switzerland}

\author[0000-0003-0960-3580]{G\'abor Worseck}
\affiliation{Institut f\"ur Physik und Astronomie, Universit\"at Potsdam, Karl-Liebknecht-Str. 24/25, 
D-14476 Potsdam, Germany}

\author[0000-0001-5758-1000]{Ricardo O. Amorín}
\affiliation{Instituto de Investigacion Multidisciplinar en Ciencia y Tecnologia, Universidad de La Serena, Raul Bitran 1305, La Serena, Chile}

\author[0000-0002-7570-0824]{Hakim Atek}
\affiliation{Institut d'astrophysique de Paris, CNRS, Sorbonne Université, 98bis Boulevrad Arago, F-75014, Paris, France}

\author[0000-0001-8587-218X]{Matthew Hayes}
\affiliation{Department of Astronomy, Oskar Klein Centre; Stockholm University; SE-106 91 Stockholm, Sweden}

\author[0000-0001-7673-2257]{Zhiyuan Ji}
\affiliation{University of Massachusetts Amherst, 710 North Pleasant Street, Amherst, MA 01003-9305, USA}

\author[0000-0002-3005-1349]{Göran Östlin}
\affiliation{Department of Astronomy, Oskar Klein Centre; Stockholm University; SE-106 91 Stockholm, Sweden}

\author[0000-0001-8419-3062]{Alberto Saldana-Lopez}
\affiliation{Department of Astronomy, University of Geneva, 51 Chemin Pegasi, 1290 Versoix, Switzerland}

\author{Trinh Thuan}
\affiliation{Astronomy Department, University of Virginia, Charlottesville, VA 22904, USA}

\correspondingauthor{Xinfeng Xu} 
\email{xinfeng@jhu.edu}


\begin{abstract}


The \mgii\ \ly\ly 2796, 2803 doublet has been suggested to be a useful indirect indicator for the escape of \lya\ and Lyman continuum (LyC) photons in local star-forming galaxies. However, studies to date have focused on small samples of galaxies with strong \mgii\ or strong LyC emission. Here we present the first study of \mgii\ probing a large dynamic range of galaxy properties, using newly obtained high signal-to-noise, moderate-resolution spectra of \mgii\ for a sample of 34 galaxies selected from the Low-redshift Lyman Continuum Survey. We show that the galaxies in our sample have \mgii\ profiles ranging from strong emission to P-Cygni profiles, and to pure absorption. We find there is a significant trend (with a possibility of spurious correlations of $\sim$ 2\%) that galaxies detected as strong LyC Emitters (LCEs) also show larger equivalent widths of \mgii\ emission, and non-LCEs tend to show evidence of more scattering and absorption features in \mgii. We then find \mgii\ strongly correlates with \lya\ in both equivalent width and escape fraction, regardless of whether the emission or absorption dominates the \mgii\ profiles. Furthermore, we present that, for galaxies categorized as \mgii\ emitters (MgE), one can adopt the information of \mgii, metallicity, and dust to estimate the escape fraction of LyC within a factor of $\sim$ 3. These findings confirm that \mgii\ lines can be used as a tool to select galaxies as LCEs and to serve as an indirect indicator for the escape of \lya\ and LyC. 


\end{abstract} 



\section{Introduction} 
\label{sec:intro}
In the last decades, considerable observational efforts have been directed towards studying the escape of Lyman continuum (LyC) photons from galaxies and attempting to explain the last phase transition of the universe, i.e., Cosmic Reionization. Various publications point out that star-forming (SF) galaxies can be responsible for the epoch of reionization (EoR).  These include studies of low-redshift galaxies \citep[z $\lesssim$ 1, e.g.,][]{Heckman01,Bergvall06, Leitet13, Borthakur14, Leitherer16,Izotov16a, Izotov16b, Puschnig17, Izotov18a,Izotov18b, Wang19, Wang21, Chisholm20, Izotov21, Chisholm22, Flury22a, Flury22b, Izotov22, Marques-Chaves22, Saldana-Lopez22, Xu22b}, and moderate- to high-redshifted galaxies \citep[z $\sim$ 2 -- 4, e.g.,][]{Robertson15, Vanzella16, deBarros16, Shapley16, Bian17, Marchi17, Marchi18, Steidel18, Vanzella18, Fletcher19, Rivera-Thorsen19, Ji20,Mestric20,Vielfaure20, Naidu22, Begley22, Rivera-Thorsen22, Marques-Chaves22b}. These galaxies are proposed to be analogs of high-redshift  (z $\sim$ 6 -- 8) ones at EoR \citep[e.g.,][]{Schaerer16, Boyett22}.

Due to the attenuation by Lyman limit systems (LLS) and/or neutral intergalactic medium (IGM), it is challenging to detect LyC at $z \gtrsim 5$ \citep{Inoue14, Worseck14, Becker21,Bosman21}. Therefore, various indirect indicators have been developed from lower redshift analogs \citep[see a summary in][]{Flury22a,Flury22b}. One of the leading indicators is the \lya\ emission \citep[e.g.,][]{Verhamme15, Henry15, Dijkstra16, Verhamme17, Jaskot19, Gazagnes20, Kakiichi21, Izotov22}. Given the resonance nature of \lya, the escape of \lya\ photons contains information about the neutral hydrogen in/around the galaxy, and leaves footprints on the \lya\ emission line profiles \citep[e.g.,][]{Izotov18b,Gazagnes20, Flury22b, LeReste22}. Nonetheless, since \lya\ photons can be also absorbed by neutral IGM, the interpretation of \lya\ profiles for high-z galaxies (z $\gtrsim$ 4) is non-trivial \citep[e.g.,][]{Stark11,Schenker14, Gronke21, Hayes21}.

The \mgii\ \ly\ly 2796, 2803 doublet has been commonly detected as a pair of absorption lines in galaxies, which trace the galactic outflows and their feedback effects \citep[e.g.,][]{Weiner09, Erb12, Finley17, WangWC22}. However, recently, \mgii\ has been found to show strong doublet emission lines in galaxies which are classified as Lyman continuum emitter (LCEs) candidates. Various publications suggest that the escape of \mgii\ correlates with that of \lya\ and LyC in SF galaxies \citep{Henry18, Chisholm20, Naidu22, Xu22b, Izotov22, Seive22}.

\mgii\ emission was studied by \cite{Henry18} in a sample of 10 compact SF galaxies. By the first time, they showed that the escape of \mgii\ correlates with the escape of \lya. This result is interpreted as evidence that both \mgii\ and \lya\ photons escape from the galaxies through a similar path of low column density gas. Indeed, \cite{Chisholm20} point out that the flux ratios between the doublet lines of \mgii\ [i.e., F(\mgii\ 2796)/F(\mgii\ 2803), hereafter, \RMg] can be used to trace the column density of neutral hydrogen. They and \cite{Xu22b} combined the \mgii\ emission, metallicity, and dust attenuation to predict \fescLyC, and found that the predicted \fescLyC\ correlates with the observed \fescLyC\ in samples of galaxies with strong \mgii\ emission lines. \cite{Xu22b} also found that galaxies selected with strong \mgii\ emission lines might be more likely to leak LyC than similar galaxies with weaker \mgii. In an independent sample, \cite{Izotov22} confirmed that escaping LyC emission is detected predominantly in galaxies with \RMg\ $\gtrsim$ 1.3, which indicates that optical depth of \mgii\ is low \citep[i.e., \protect$\tau_{2803} $$\lesssim$ 0.5,][]{Chisholm20}. Therefore, a high \RMg\ ratio can be used to select LCEs candidates. 

\mgii\ emission lines are also detected in higher redshift SF galaxies. For example, in the stacks of bright \lya\ emitters (LAEs) at z $\sim$ 2, \cite{Naidu22} found that LCE candidates tend to have \RMg\ close to 2, and the \mgii\ emission is closer to the systemic velocity (instead of redshifted in non-LCEs). In addition, \cite{Witstok21} found in a lensed z $\sim$ 5 galaxy, the escape of \mgii\ photons is consistent with that of \lya. However, \cite{Katz22} pointed out from the hydro-cosmological simulations that \mgii\ is a useful diagnostic of \fescLyC\ only in the optically thin regime.


Though all of these studies support the hypothesis that strong \mgii\ emission lines and a high \RMg\ ratio can serve as a good indirect indicator for \lya\ and LyC escape, there exist two caveats. (1) Existing samples are commonly small and only focused on SF galaxies with strong \mgii\ emission lines and/or galaxies as strong LyC emitters. The latter also results in substantial \mgii\ emission lines. Similar SF galaxies with \mgii\ as weaker emission lines, P-Cygni profiles, or absorption lines have not been systematically studied in observations. (2) Some of these studies only have low signal-to-noise (S/N) \mgii\ spectra \citep[e.g.,][]{Naidu22, Xu22b, Izotov22}. Their determinations of \RMg, and the subsequent inference of \mgii\ optical depth, have more scatter due to the low S/N spectra.


In this paper, we further study \mgii\ as an indirect indicator for the escape of \lya\ and LyC, while we attempt to mitigate the two caveats noted above. We focus on galaxies from Low-redshift Lyman Continuum Survey \citep[LzLCS, ][]{Flury22a}, where a large sample of 66 LCEs candidates are selected without reference to their \mgii\ emission lines. For 34 out of 66 LzLCS galaxies, we present ground-based follow-up spectroscopy of the \mgii\ feature. These data provide higher spectral resolution and S/N than SDSS/BOSS, along with coverage of \mgii\ in galaxies where it is blueward of the SDSS bandpass. As we will show, these data achieve more secure measurements of the \mgii\ properties, ultimately validating the use of  \mgii\ as an indirect indicator for the escape of \lya\ and LyC.




The structure of the paper is as follows. In Section \ref{sec:obs},we introduce the observations, data reduction, and basic measurements of optical emission lines. In Section \ref{sec:analyses}, we present how we derive various important parameters from the observed \mgii\ doublet. We show the main results in Section \ref{sec:results}, including the comparisons of \mgii\ with \lya\ and LyC. We conclude the paper in Section \ref{sec:Conclusion}.


\section{Observations, Data Reductions, and Basic Analyses}
\label{sec:obs}

\subsection{Ultraviolet Spectra for LyC and \lya\ Regions}

In this study, we focus on LCE candidates from the Low-redshift Lyman Continuum Survey \citep[LzLCS,][]{Flury22a}, which contains 66 SF galaxies at z $\sim$ 0.3. The survey obtained rest-frame ultraviolet (UV) spectra for each galaxy with the G140L grating on \textit{Hubble Space Telescope}/\textit{Cosmic Origins Spectrograph} (HST/COS) under program GO 15626 (PI: Jaskot). Both LyC and \lya\ regions have been covered. The detailed data reductions of G140L data as well as the analysis of \lya\ and LyC escape are presented in \cite{Flury22a}. In this paper, we adopt their derived quantities for LyC and \lya. These mainly include their escape fractions and the equivalent widths (EW) for \lya.

\begin{table*}
	\centering
	\caption{Follow-up Observations and Basic Properties for Galaxies in Our Sample}
	\label{tab:obs}
	\begin{tabular}{lllcllcccc} 
		\hline
		\hline
		ID & RA & Dec 	    & $z^{1}$   & Instrument$^{2}$ &	Date$^{3}$ &	Exp.$^{3}$  & SDSS-u$^{4}$ & $E(B-V)_\text{MW}^{5}$  &  $E(B-V)_\text{int.}^{6}$  \\
		\hline
		& &         	&          &  & (mm/dd/yyyy)         & (s) & (mag) &  & \\
		\hline
        J0957+2357  & 09:57:00 & +23:57:09      &0.2444 & MMT/Blue	    & 04/08/2019	& 	4800  &18.42  & 0.0287 & 0.3007 \\
        J1314+1048  & 13:14:19 & +10:47:39      &0.2960 & MMT/Blue	    & 04/08/2019	& 	3600  &19.69  & 0.0371 & 0.1621 \\
        J1327+4218  & 13:26:33 & +42:18:24      &0.3176 & MMT/Blue	    & 04/08/2019	& 	3600  &20.48  & 0.0173 & 0.1641 \\
        J1346+1129  & 13:45:59 & +11:28:48      &0.2371 & MMT/Blue	    & 04/08/2019	& 	3600  &19.00  & 0.0231 & 0.2022 \\
        J1410+4345  & 14:10:13 & +43:44:35      &0.3557 & MMT/Blue	    & 04/08/2019	& 	7200  &21.70  & 0.0196 & 0.1413 \\

        J0926+3957  & 09:25:52 & +39:57:14      &0.3141 & MMT/Blue	    & 04/09/2019	& 	7200  &21.27  & 0.0270 & 0.1364 \\
        J1130+4935  & 11:29:33 & +49:35:25      &0.3448 & MMT/Blue	    & 04/09/2019	& 	7200  &21.50  & 0.0292 & 0.0446 \\
        J1133+4514  & 11:33:04 & +65:13:41      &0.2414 & MMT/Blue	    & 04/09/2019	& 	7200  &20.14  & 0.0158 & 0.0886 \\
        J1246+4449  & 12:46:19 & +44:49:02      &0.3220 & MMT/Blue	    & 04/09/2019	& 	4800  &20.48  & 0.0200 & 0.1595 \\

        J0723+4146  & 07:23:26 & +41:46:08      &0.2966 & MMT/Blue	    & 02/19/2020	& 	7200  &20.89  & 0.0467 & 0.0097 \\
        J0811+4141  & 08:11:12 & +41:41:46      &0.3329 & MMT/Blue	    & 02/19/2020	& 	7200  &21.17  & 0.0383 & 0.1150 \\
        J1235+0635  & 12:35:19 & +06:35:56      &0.3326 & MMT/Blue	    & 02/19/2020	& 	6000  &20.72  & 0.0333 & 0.0782 \\

        J0814+2114  & 08:14:09 & +21:14:59      &0.2271 & MMT/Blue	    & 02/20/2020	& 	1800  &18.91  & 0.0336 & 0.1800 \\
        J0912+5050  & 09:12:08 & +50:50:09      &0.3275 & MMT/Blue	    & 02/20/2020	& 	9600  &20.56  & 0.0245 & 0.1088 \\
        J1301+5104  & 13:01:28 & +51:04:51      &0.3476 & MMT/Blue	    & 02/20/2020	& 	2500  &20.04  & 0.0222 & 0.0974 \\

        J0047+0154  & 00:47:43 & +01:54:40      &0.3535 & MMT/Blue	    & 01/09/2021	& 	3600  &20.28  & 0.0312 & 0.1760 \\
        J0826+1820  & 08:26:52 & +18:20:52      &0.2972 & MMT/Blue	    & 01/09/2021	& 	7200  &21.35  & 0.0328 & 0.0300 \\
        J1158+3125  & 11:58:55 & +31:25:59      &0.2430 & MMT/Blue	    & 01/09/2021	& 	2700  &19.27  & 0.0226 & 0.1037 \\
        J1248+1234  & 12:48:35 & +12:34:03      &0.2635 & MMT/Blue	    & 01/09/2021	& 	6900  &20.16  & 0.0519 & 0.0627 \\

        J0113+0002  & 01:13:09 & +00:02:23      &0.3060 & MMT/Blue	    & 01/10/2021	& 	7200  &20.56  & 0.0312 & $<$1E-4 \\
        J0129+1459  & 01:29:10 & +14:59:35      &0.2799 & MMT/Blue	    & 01/10/2021	& 	4800  &20.11  & 0.0688 & 0.0729 \\
        J0917+3152  & 09:17:03 & +31:52:21      &0.3003 & MMT/Blue	    & 01/10/2021	& 	3300  &19.78  & 0.0204 & 0.1920 \\
        J1033+6353  & 10:33:44 & +63:53:17      &0.3465 & MMT/Blue	    & 01/10/2021	& 	2700  &19.84  & 0.0160 & 0.0819 \\
        J1038+4527  & 10:38:16 & +45:27:18      &0.3256 & MMT/Blue	    & 01/10/2021	& 	2700  &19.40  & 0.0241 & 0.2506 \\
        \hline

        J0036+0033  & 00:36:01 & +00:33:07         &0.3480 & VLT/X-Shooter	    & 11/05/2020	& 	2800  &21.83  & 0.0278 & 0.2007 \\
        J0047+0154  & 00:47:43 & +01:54:40         &0.3537 & VLT/X-Shooter	    & 10/23/2020	& 	5500  &20.28  & 0.0312 & 0.0699 \\
        J0113+0002  & 01:13:09 & +00:02:23         &0.3062 & VLT/X-Shooter	    & 10/23/2020	& 	5500  &20.56  & 0.0312 & 0.1144 \\
        J0122+0520  & 01:22:17 & +05:20:44         &0.3655 & VLT/X-Shooter	    & 10/23/2020	& 	5500  &21.38  & 0.0559 & 0.1166 \\
        J0814+2114  & 08:14:09 & +21:14:59         &0.2271 & VLT/X-Shooter	    & 12/21/2020	& 	2800  &18.91  & 0.0335 & 0.1984 \\
        J0911+1831  & 09:11:13 & +18:31:08         &0.2622 & VLT/X-Shooter	    & 01/16/2021	& 	2800  &19.79  & 0.0279 & 0.1106 \\
        J0958+2025  & 09:58:38 & +20:25:08         &0.3016 & VLT/X-Shooter	    & 02/04/2021	& 	2800  &19.86  & 0.0347 & 0.0650 \\
        J1310+2148  & 13:10:37 & +21:48:17         &0.2830 & VLT/X-Shooter	    & 04/10/2021	& 	5500  &20.40  & 0.0204 & 0.0924 \\
        J1235+0635  & 12:35:19 & +06:35:56         &0.3327 & VLT/X-Shooter	    & 01/12/2022	& 	5500  &20.72  & 0.0333 & 0.0782 \\
        J1244+0215  & 12:44:23 & +02:15:40         &0.2394 & VLT/X-Shooter	    & 03/08/2022	& 	2800  &19.56  & 0.0424 & 0.0673 \\
        \hline
        J0834+4805  & 08:34:40 & +48:05:41         &0.3425 & HET/LRS2	        & 12/30/2021	& 	5400  &20.62  & 0.0383 & 0.1741 \\
        J0940+5932  & 09:40:01 & +59:32:44         &0.3716 & HET/LRS2	        & 01/24/2022	& 	5400  &20.80  & 0.0380 & 0.4158 \\
        J1517+3705  & 15:17:07 & +37:05:12         &0.3533 & HET/LRS2	        & 07/22/2022	& 	6300  &20.87  & 0.0411 & 0.0005 \\
        J1648+4957  & 16:48:49 & +49:57:51         &0.3818 & HET/LRS2	        & 05/27/2022	& 	5400  &21.93  & 0.0374 & 0.0070 \\


		\hline
		\hline
	\multicolumn{10}{l}{%
  	\begin{minipage}{17cm}%
	Note. --\\
    	(1)\ \ Redshift of the objects derived from fitting the Balmer emission lines.\\
    	(2)\ \ Instruments that are used for the follow-up observations (Section \ref{sec:optical}).\\
    	(3)\ \ Observation start-date and exposure time in seconds, respectively.\\
    	(4)\ \ The u-band magnitudes from SDSS photometry.\\
    	(5)\ \ Milky Way dust extinction obtained from Galactic Dust Reddening and Extinction Map \citep{Schlafly11} at NASA/IPAC Infrared Science Archive.\\
    	(6)\ \ The internal nebular dust extinction of the galaxy derived from Balmer lines (Section \ref{sec:Basic}).\\
    	
  	\end{minipage}%
	}\\
	\end{tabular}
	\\ [0mm]
	
\end{table*}


\begin{figure*}
\center
	\includegraphics[angle=0,trim={0.2cm 1.75cm 0.0cm 8.5cm},clip=true,width=0.99\linewidth,keepaspectratio]{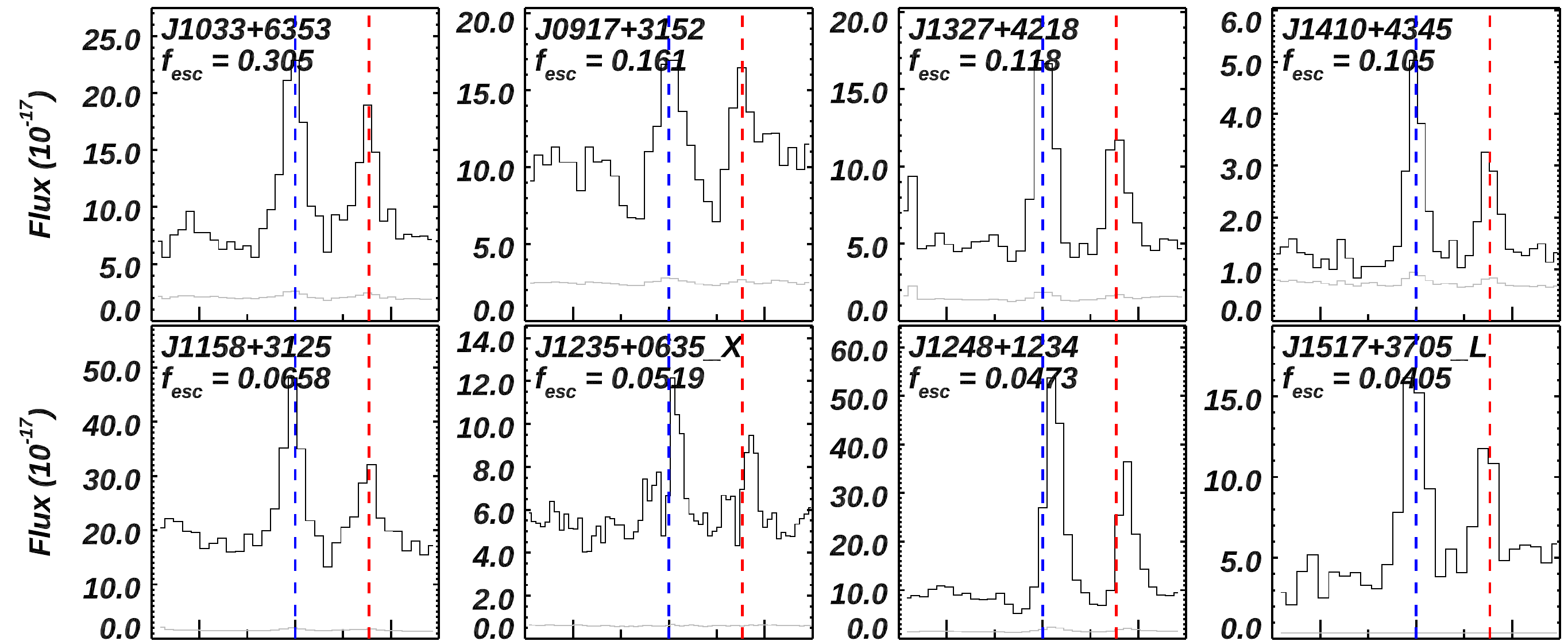}
	
	\includegraphics[angle=0,trim={0.0cm 0.1cm 0.0cm 8.6cm},clip=true,width=1\linewidth,keepaspectratio]{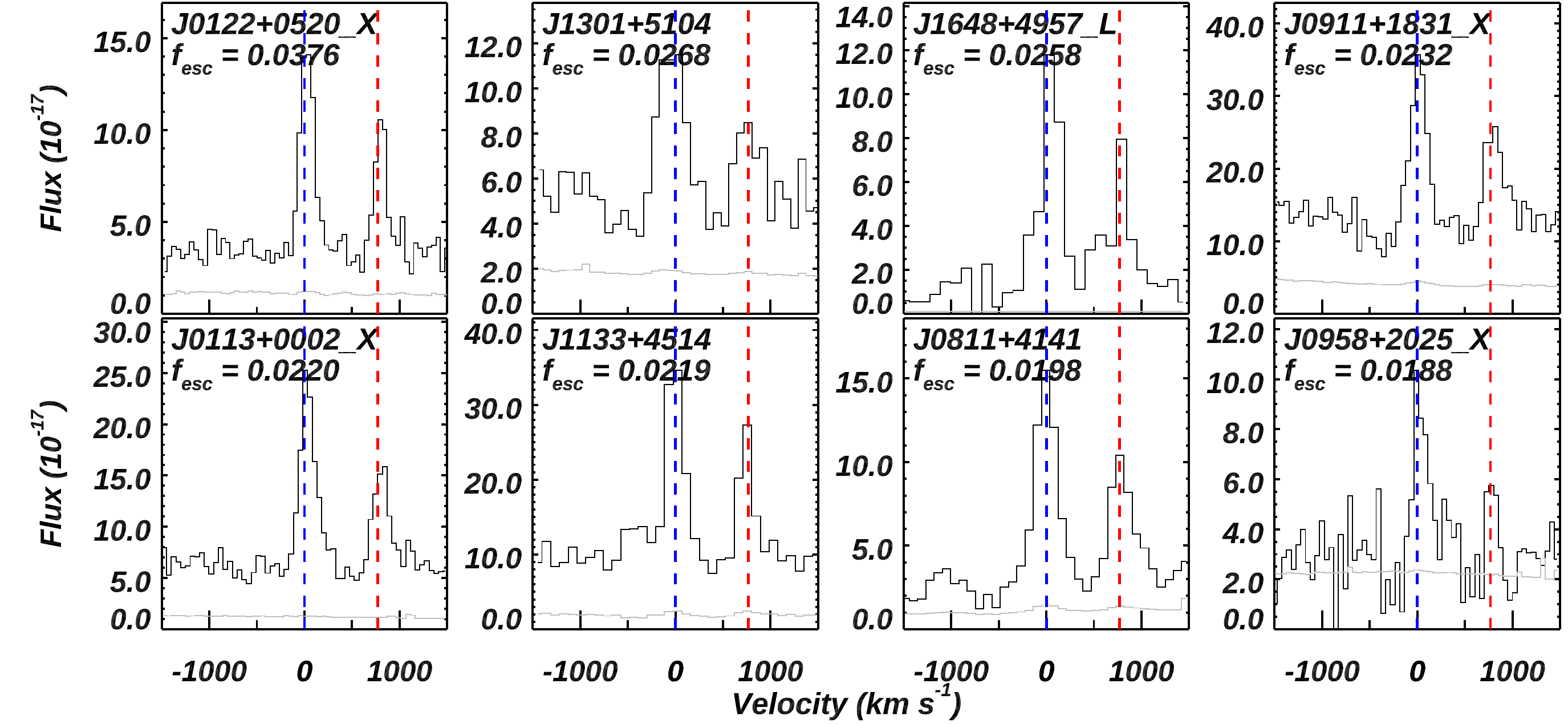}

\caption{\normalfont{The final reduced \mgii\ spectra for LzLCS galaxies in velocity space, with data taken from either MMT blue channel spectrograph or VLT/X-Shooter spectrograph. For X-Shooter and LRS2 observations, we mark them with an extra `X' and `L' at the end of object names, respectively. The y-axes are in units of 10$^{-17}$ ergs s$^{-1}$ cm$^{-2}$ \AA$^{-1}$. The data and corresponding errors are shown in black and gray. Objects are ordered by measured \fescLyC\ values published in \cite{Flury22a}, which are also shown in the top-left corner of each panel. The blue and red lines represent the position of $v$ = 0 km s$^{-1}$ for \mgii\ \ly 2796 and 2803, respectively.  } }
\label{fig:MgII-1}
\end{figure*}

\begin{figure*}
\center
	\includegraphics[angle=0,trim={0.2cm 1.75cm 0.0cm 8.5cm},clip=true,width=0.99\linewidth,keepaspectratio]{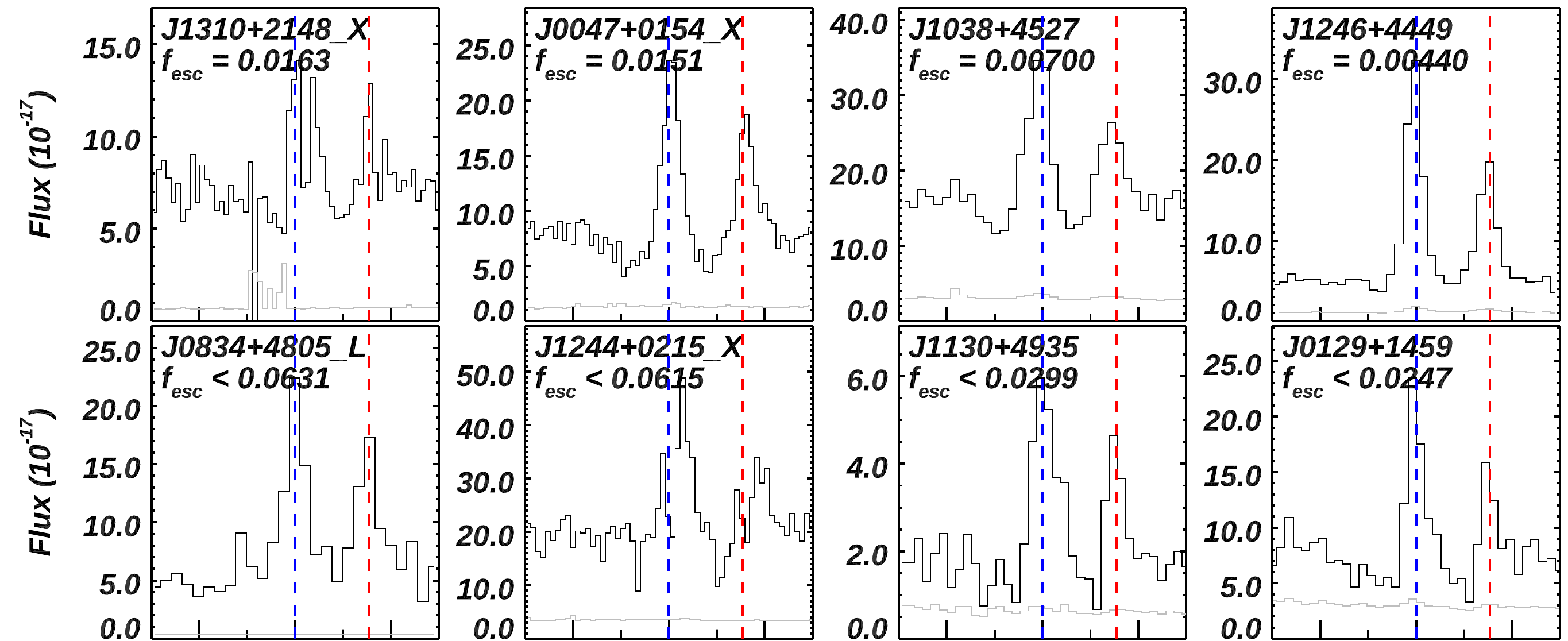}
	
	\includegraphics[angle=0,trim={0.0cm 1.75cm 0.0cm 8.6cm},clip=true,width=1\linewidth,keepaspectratio]{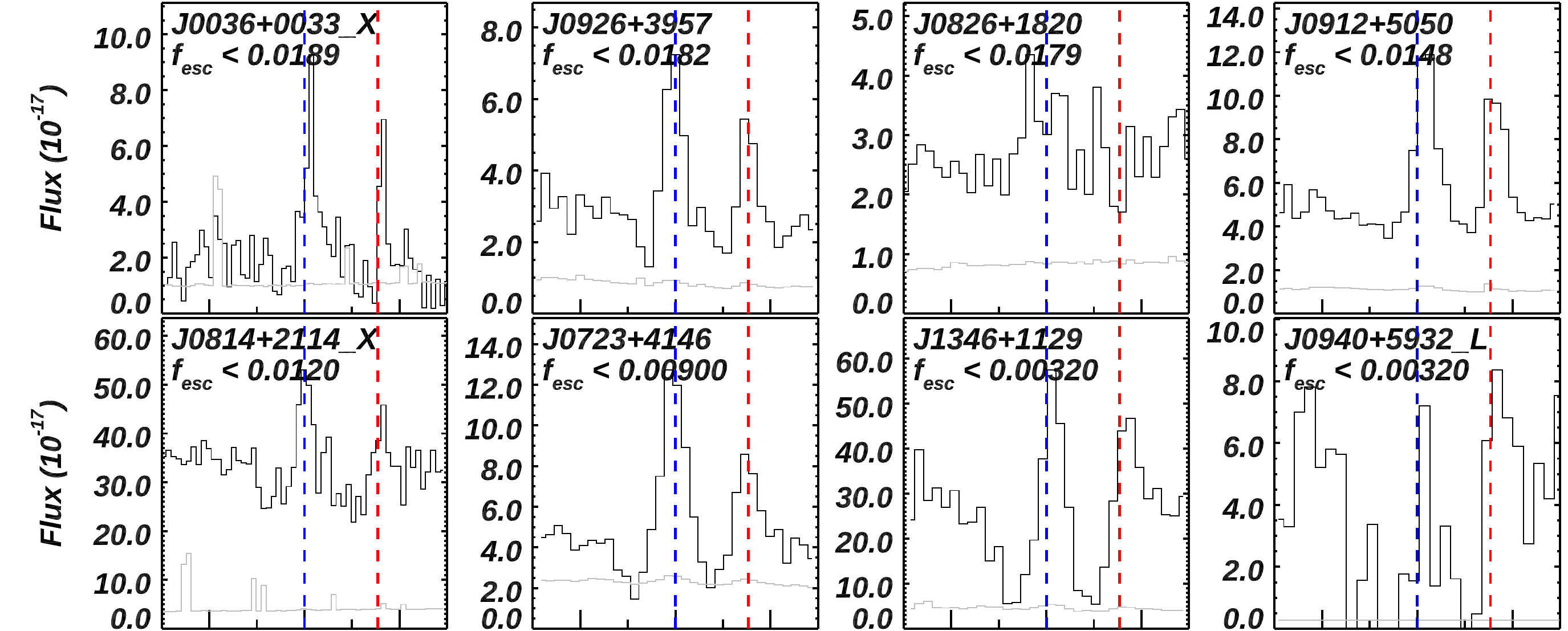}

	\includegraphics[angle=0,trim={0.0cm 5.4cm 0.0cm 8.6cm},clip=true,width=1\linewidth,keepaspectratio]{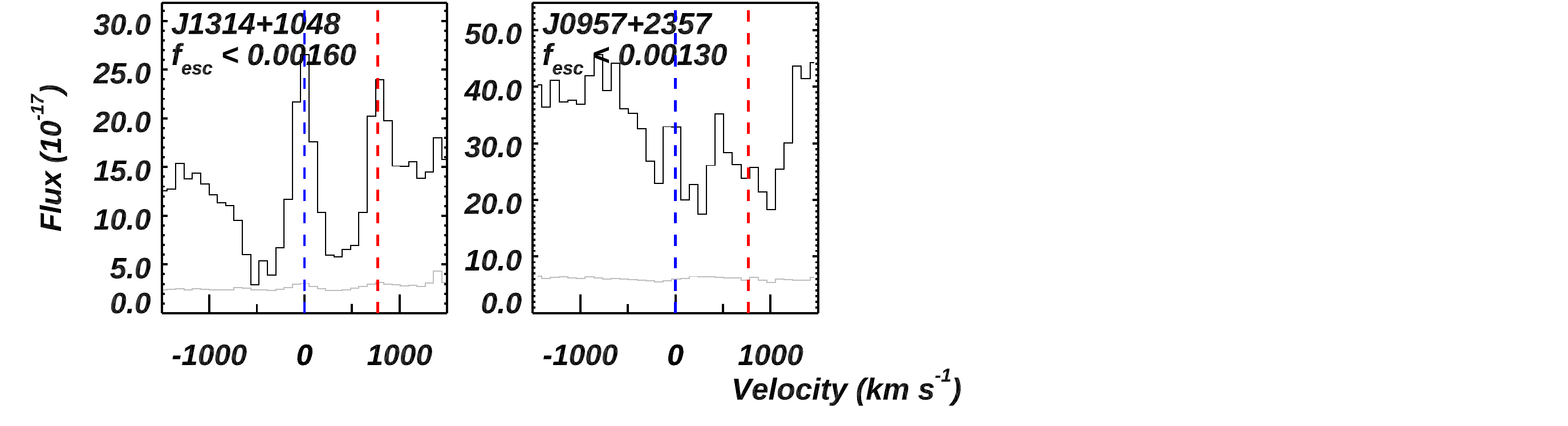}
	
\caption{\normalfont{Same as Figure \ref{fig:MgII-1} but for galaxies with lower \fescLyC. Upper limits of \fescLyC\ are presented for galaxies that have non-detections of LyC flux \citep{Flury22a}. From LCEs to non-LCEs, \mgii\ line profiles show a clear transition from strong emission lines to P-Cygni profiles to strong absorption lines. See more discussion in Section \ref{sec:trends}.} }
\label{fig:MgII-2}
\end{figure*}

\subsection{Optical Spectra for \mgii\ Regions}
\label{sec:optical}
The galaxies in the LzLCS sample already have SDSS or BOSS spectra, where the blue wavelength coverages end at 3800\angstrom\ and 3650\angstrom, respectively. Thus, the \mgii\ features are observed by SDSS or BOSS only for galaxies with z $\gtrsim$ 0.35 or 0.30, respectively. Even for the cases with existing \mgii\ spectra from SDSS/BOSS, as discussed in \cite{Xu22b}, the S/N of the intrinsically weak \mgii\ lines and the spectral resolution ($\sim$ 1500) are commonly too low. In this case, the measurements of \mgii, especially \RMg, have large error bars. Therefore, we have obtained higher S/N and higher spectral resolution observations for 34 galaxies from the LzLCS sample. These galaxies are observed by either the Multiple Mirror Telescope (MMT) or the Very Large Telescope (VLT), or the Hobby-Eberly Telescope (HET). In this paper, we focus on studying the \mgii\ properties from this subsample of 34 galaxies. The observation details and data reductions are listed in Table \ref{tab:obs} and discussed below.

\subsubsection{MMT observations and Data Reductions}
A total of 24 galaxies from the LzLCS sample have been observed by MMT. We adopt the blue channel spectrograph using a 1\arcsec\ slit with the 832 lines/mm grating at the second order. This leads to a spectral resolution of $\sim$ 1\angstrom\ ($\sim$ 90 km s$^{-1}$ near the \mgii\ region). The observations were conducted on 6 nights in 3 different semesters (2019A, 2020A, 2021A). The exposure time is between 30 mins to 180 mins, depending on the brightness of the target (Table \ref{tab:obs}). We stay towards lower airmass ($\lesssim$ 1.3) and, for every exposure, we reset the slit at the parallactic angle. We reduce the data following the methodology described in \cite{Henry18} using IDL + IRAF routines. The wavelength calibration is applied from the HeArHgCd arc lamps. By matching the arc lines, we find the root-mean-square of the residuals is $<$~0.1\angstrom\ ($\sim$ 10 km $^{-1}$ around \mgii\ spectral regions).


Given the short wavelength coverage of the blue channel spectrograph in MMT ($\sim$ 3100 -- 4100\angstrom), the only major line covered is the \mgii\ doublet. Therefore, we also adopt SDSS/BOSS spectra for measuring other optical lines (Section \ref{sec:Basic}). For each galaxy, we calculate its u-band magnitude from the MMT spectra and scale it to the galaxy's u-band magnitude from the SDSS photometry. This accounts for any slit losses between MMT and SDSS observations.





\subsubsection{VLT observations and Data Reductions}

We also include 10 LzLCS sources observed by the X-Shooter spectrograph mounted on VLT as part of the ESO program ID 106.215K.001 (PI: Schaerer). Observations were carried out between Fall 2020 and Spring 2022. We use 1.0\arcsec, 0.9\arcsec, and 0.9\arcsec\ slits in the UVB, VIS and NIR arms providing resolution power of $\sim$ 5400, 8900, and 5600, respectively. This yields a spectral resolution of $\sim$ 50 km s$^{-1}$ near the \mgii\ regions. Observations were performed in nodding-on-slit mode with a standard ABBA sequence and total on-source exposure times of 46 mins or 92 min, depending on the brightness of each source (Table \ref{tab:obs}). We reduce X-Shooter data following the methods in \cite{Marques-Chaves22} adopting the standard ESO Reflex reduction pipeline \citep[version 2.11.5,][]{Freudling13}.

For each galaxy, we also calculate the u-band magnitude from the VLT spectra and match it to the galaxy's u-band magnitude from the SDSS photometry. Four of our galaxies are observed by both MMT and VLT. We have checked that the \mgii\ spectral profiles from the two telescopes are similar, and the \mgii\ line flux ratio (i.e., R) are consistent within errors. This is expected since galaxies in our sample are rather compact (with UV half-light-radius $\lesssim$ 0.4\arcsec) and are smaller than the slit sizes. We finally adopt these galaxies' VLT observations in our analyses, given their higher S/N.

\subsubsection{HET observations and Data Reductions}
We include 4 additional galaxies from the LzLCS sample, which are observed by the Low-Resolution Spectrograph (LRS2) on the Hobby-Eberly Telescope \citep{Ramsey98}. LRS2 is an integral field spectrograph with nearly complete spatial sampling, and a native spatial scale of 0.25\arcsec\ $\times$ 0.25\arcsec\ spaxels with an average of 1.25\arcsec\ seeing \citep[][]{Chonis16}. LRS2 has a wavelength coverage from 3600\angstrom\ to 10,000\angstrom, and its spectral resolution around \mgii\ region is 1.63\angstrom. To match our MMT and VLT slit sizes, we extract the \mgii\ spectra in the central 1.0\arcsec\ $\times$ 1.0\arcsec\ aperture. We reduce the LRS2 data using the same methods in \cite{Seive22}, where we adopt the HET LRS2 pipeline, Panacea\footnote{https://github.com/grzeimann/Panacea}, to
perform the initial reductions, including fiber extraction, wavelength, calibration, astrometry, and flux calibration. For each galaxy, we also calculate the u-band magnitude from the LRS2 spectra and match it to the galaxy's u-band magnitude from the SDSS photometry.


\subsubsection{Summary of Optical Spectra}
Overall, we obtained higher-quality data for 34 out of 66 galaxies from the LzLCS sample. We show the final reduced \mgii\ spectra for these galaxies in Figures \ref{fig:MgII-1} and \ref{fig:MgII-2}, and have ordered them by decreasing absolute escape fraction of LyC (\fescLyC) measured from fitting the UV continuum \citep[reported in][]{Flury22a}. Based on their LyC measurements, these galaxies have \fescLyC\ range between 0 and 30\%. Of the 34 galaxies, 20 are classified as Lyman continuum emitters (LCEs, sometimes referred as LyC ``leakers''), which have LyC flux detected with 97.725\% confidence \citep{Flury22a}. The other 12 galaxies are classified as non-LCEs. We show the derived \fescLyC\ values at the top-left corners of each panel, while we present \fescLyC\ upper limits for non-LCEs. We mark the galaxies observed by X-Shooter or LRS2 with an extra `X' or `L', respectively, at the end of their object names in Figures \ref{fig:MgII-1} and \ref{fig:MgII-2}.



\subsection{Measurements of Optical Emission Lines}
\label{sec:Basic}

For galaxies that have new optical spectra as described above, we measure several optical emission lines whenever covered, including \mgii, [\oii], [\oiii], and Balmer lines. For each galaxy, we first correct the spectra for Milky Way extinction using the Galactic Dust Reddening and Extinction Map \citep{Schlafly11} at NASA/IPAC Infrared Science Archive, assuming the extinction law from \cite{Cardelli89}. The redshift of the galaxy is matched to the peak of Balmer emission lines.

We determine the continuum flux for the \mgii\ spectral region by adopting a linear fit to the spectra $\sim$ $\pm$ 2000 km~s$^{-1}$ around the systemic velocity. Then we split the spectra at the midpoint between the two lines, i.e., 2799.1\angstrom, to represent the spectral regions for 2796 and 2803, separately. For each \mgii\ line, we also split it into the absorption (below the continuum) and emission (above the continuum) parts. After that, we integrate the separate spectral regions to get the flux and EW. The corresponding errors on these quantities are estimated through a Monte Carlo (MC) simulation where we perturb the spectrum 10$^{4}$ times according to the observed 1$\sigma$ uncertainties. These values are reported in Table \ref{tab:MgII}. Note that we do not correct the \mgii\ line fluxes by internal dust extinction of the galaxy. This is because \mgii\ photons are resonantly scattered like \lya\ and robust correction is difficult \citep[][]{Henry18, Chisholm20, Xu22b}.

For other optical lines, we measure their flux and EW similarly as \mgii. However, unlike \mgii, since they are not resonant lines, we also correct the spectra by the internal dust extinction for the galaxy before the measurements. The internal dust extinction ($E(B-V)_\text{int.}$) for each galaxy is measured from Balmer lines following the methods in \cite{Xu22b}. For galaxies that only have new MMT observations, since the MMT blue channel does not cover the Balmer lines, we adopt $E(B-V)_\text{int.}$ derived in \cite{Flury22a} based on their SDSS spectra. The final $E(B-V)_\text{int.}$ values are reported in the last column in Table \ref{tab:obs}.

\begin{figure}
\center

	\includegraphics[angle=0,trim={0.0cm 0.8cm 0.0cm 0.0cm},clip=true,width=1.0\linewidth,keepaspectratio]{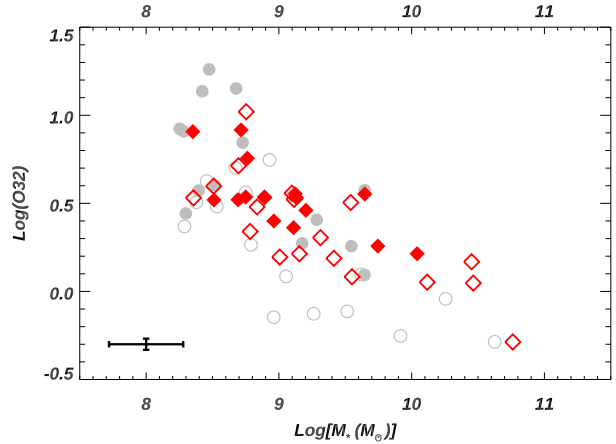}%

\caption{\normalfont{Comparisons of the sub-sample studied in this paper (red) to the other galaxies in the LzLCS parent sample (gray, see Section \ref{sec:obs}). Gray-filled and open symbols stand for galaxies from LzLCS, which are classified as LyC emitter and non-emitters, respectively \citep{Flury22a}. Red-filled and open symbols represent galaxies that are \mgii\ emitter and non-emitters, respectively (see definitions in Section \ref{sec:geometry}). The mean 1$\sigma$ error bars are shown at the bottom-left of the panel.    } }
\label{fig:CompLzLCS}
\end{figure}

In Figure \ref{fig:CompLzLCS}, we compare the sub-sample adopted in this paper (red) to the rest of the galaxies in LzLCS (gray). We show two general observables that can be measured at high-redshift, including O32 = flux ratio of [\oiii] \ly 5007/[\oiii] \ly 3727 and stellar mass derived from spectral energy distribution (SED) fitting reported in \cite{Flury22a}. Our sub-sample of galaxies is randomly selected from the LzLCS parent sample to ensure a large dynamic range in galaxy properties.

\section{Analyses}
\label{sec:analyses}

In this section, we present the methodology to derive important properties from \mgii\ lines. We first discuss the significant trends in \mgii\ line profiles in Section \ref{sec:trends}. Then we show the plausible geometry for \mgii\ photon escape in Section \ref{sec:geometry}. We present the methods to derive the \mgii\ escape fractions (\fescMgII) from photoionization models in Section \ref{sec:fescMgII}. Finally, we discuss how to predict the escape fraction of LyC (\fescLyCPred) from \mgii, metallicity, and dust attenuation in Section \ref{sec:fescLyC}.





\subsection{Significant Trends of \mgii\ Line Profiles}
\label{sec:trends}


In Figures \ref{fig:MgII-1} and \ref{fig:MgII-2}, we show the \mgii\ spectra from our galaxies in the order of decreasing \fescLyC\ derived from HST/COS G140L spectra \citep{Flury22a}. There exist a significant trend that galaxies detected as strong LCEs also show strong \mgii\ emission lines, and non-LCEs present more absorption features in \mgii. We apply the Kendall $\tau$ test between EW(\mgii) and \fescLyC, where we have considered the upper limits following \cite{Akritas96}. This leads to the probability of a spurious correlation, $p$ = 0.0216, which confirms the strong trend. The former half of this trend is consistent with previous observations of strong LCEs \citep{Izotov22, Xu22b}. Nonetheless, our sample is the first to show that this trend indeed extends to non-LCEs. This can be explained as LCEs have more optically thin clouds in/around the galaxy than non-LCEs, so both the \mgii\ and LyC photons can escape with less absorption and scattering \citep{Chisholm22}. This is also consistent with the expectations from simulations \citep{Katz22}.


Notably, a high \fescLyC\ (= 16.1\%) was measured for galaxy J0917+3152, but its \mgii\ profiles also have clear absorption features. This can be explained by the high metallicity of J0917+3152, i.e., 12+log(O/H) = 8.46, which is the highest in our sample. Thus, for this object, there exist more magnesium atoms given the same amount of hydrogen atoms. In this scenario, the clouds around J0917+3152 become optically thick to \mgii\ when it is still optically thin to LyC photons. Overall, the significant trend for \mgii\ profiles from LCEs to non-LCEs is valid for galaxies with lower metallicity (in our case, 12+log(O/H) $<$ 8.4). In these galaxies, the surrounding gas/clouds become optically thick to \mgii\ and LyC photons at similar depths \citep{Chisholm20}.

\subsection{Possible Geometry for the Escape of \mgii\ photons and Constraints on Models}
\label{sec:geometry}
The escape of \mgii\ photons is first discussed in details in \cite{Chisholm20} (hereafter, the Chisholm model, see their Section 6.4). This model assumes that \mgii\ photons escape through a partial coverage geometry or sometimes referred as the picket-fence geometry \citep[see also][]{Gazagnes18, Chisholm18, Saldana-Lopez22, Xu22b}: 

\begin{equation}\label{eq:PC}
\begin{aligned} 
    f_\text{esc}^\text{Mg II} = \frac{F_\text{obs}}{F_\text{int}} =  C_{f}(\text{Mg II})e^{-\tau_\text{thick}} + [1-C_{f}(\text{Mg II})]e^{-\tau_\text{thin}}
\end{aligned} 
\end{equation}
where F$_\text{obs}$ and F$_\text{int}$ are the observed and intrinsic flux of \mgii, respectively; \CF\ is the covering fractions for the optically thick paths of \mgii; and \Tthick\ and \Tthin\ are the optical depths for \mgii\ at optically thick and thin paths, respectively. In the optically thick paths, it is usually assumed that \Tthick\ $\gg$ 1 such that no \mgii\ photons are observed through this path. In this model, \cite{Chisholm20} also found:

\begin{equation}\label{eq:MgIItau}
\begin{aligned} 
    R = \frac{F_\text{2796,obs}}{F_\text{2803,obs}} = 2e^{-\tau_{2803,thin}}
\end{aligned} 
\end{equation}
where \RMg\ is the emission line flux ratio between the \mgii\ doublet. This model has proven to be successful in \cite{Chisholm20} and \cite{Xu22b} for galaxies with strong \mgii\ emissions, where \mgii\ photons escape from the galaxy through mostly optically thin paths. Furthermore, \cite{Katz22} have tested this model in their hydro-cosmological simulations for EoR galaxy analogs. They find the actual line-of-sight (LOS) \fescMgII\ match well with the predicted ones from the Chisholm model for galaxies with low metallicity (thus less dusty) and high \fescLyC. These galaxies have \mgii\ line profiles dominated by emission. 

However, as described in Section \ref{sec:trends}, given the large dynamic ranges of our sample by design, our galaxies have \mgii\ profiles ranging from strong emission to P-Cygni profiles and pure absorption. In the latter two cases, at least two factors complicate the applications of the Chisholm model. (1) The measurements of \RMg\ from the spectra are not well-defined due to the absorption in \mgii\ profiles. Thus, the derived $\tau_{2803,thin}$ from \RMg\ has large uncertainties. (2) \mgii\ doublet are resonant lines. Thus, the effect of dust for \mgii\ is more substantial, which can cause strong absorption and scattering features in the spectra (e.g., J0957+2357). However, this effect cannot be described in simple terms, thus, is not included in the Chisholm model. Similarly, \cite{Katz22} comment that the Chisholm model is likely inadequate to predict \fescMgII\ for metal-rich (thus more dusty) galaxies in their simulations.



Currently, in this paper, to highlight the limitations of the Chisholm model, we manually split galaxies in our sample into two categories in our analyses. Galaxies with \mgii\ as strong emission, minimal absorption, and symmetric line profiles are categorized as \MgE\ (acronym for \mgii\ emitter), while the others belong to \nMgE. This subjective classification is similar to what was adopted in \cite{Katz22}. The Chisholm model should apply well to the former since \mgii\ photons suffer little resonant scattering effects, but not perfectly to the latter. We show the category of each galaxy in the second to last column in Table \ref{tab:MgII}. We discuss further how we handle these two categories in Section \ref{sec:LyC}.





\subsection{The Method to Estimate the Escape Fraction of \mgii}
\label{sec:fescMgII}


\cite{Henry18} have first introduced that one can derive the intrinsic flux of \mgii\ from a correlation between \mgii/[\oiii] and [\oiii]/[\oii] (hereafter, the Henry model):

\begin{equation}\label{eq:MgII-O32_part1}
\begin{aligned} 
    & R_{2796} & = A_{2} \times O_{32}^2 + A_{1} \times O_{32} +A_{0} 
\end{aligned}    
\end{equation}

\begin{equation}\label{eq:MgII-O32_part2}
\begin{aligned} 
    & R_{2796}  &= log(F_\text{int}(\text{Mg II } \lambda2796)/F_\text{int}(\text{[O III] } \lambda5007))\\
    & O_{32}       &= log(F_\text{int}(\text{[O III]}\ \lambda5007)/F_\text{int}(\text{[O II] } \lambda3727))
\end{aligned}  
\end{equation}
where the emission line fluxes are all intrinsic, i.e., before the attenuation by dust and absorption in the LOS. Hereafter, we use $F_\text{int}$ to denote the intrinsic flux. A$_{0}$, A$_{1}$, and A$_{2}$ are coefficients that are dependent on the gas phase metallicity of the galaxy, but little on the ionization parameters and spectral slopes \citep{Henry18}. Combining $F_\text{int}$(\mgii) with the measured $F_\text{obs}$(\mgii) from the spectra, one can derive \fescMgII.

The photoionization models in \cite{Henry18} considers ionization bounded (IB) geometry, where most of the cloud remains neutral and is optically thick to escaping photons. However, LCEs with strong \mgii\ emission lines can be partly density bounded (DB, i.e., mostly optically thin) given their high O$_{32}$ values observed \citep[e.g.,][]{Izotov16a, Izotov16b, Izotov18a, Izotov18b, Izotov21,Flury22a, Flury22b, Xu22b}. Thus, \cite{Xu22b} update the correlation coefficients in the Henry model to take into account the DB scenario. At a fixed metallicity, they also find the different models from DB and IB only move the correlation along the line defined in Equation (\ref{eq:MgII-O32_part1}). This should explain why \cite{Katz22} find the Henry model is a relatively good match to galaxies in their simulations, but with moderate scatter given the different metallicities of their simulated galaxies.

Given the derived \fescMgII, one can solve \CF\ and $\tau_\text{thin}$ from Equations (\ref{eq:PC}) and (\ref{eq:MgIItau}). Note this requires robust measurements of \mgii\ doublet flux ratio, i.e., $R$ in Equation (\ref{eq:MgIItau}). As discussed in Section \ref{sec:geometry}, this can be achieved in \MgE, but hardly in \nMgE\ due to the absorption features in \mgii. Since \CF\ and $\tau_\text{thin}$ are then adopted to predict \fescLyC, accurately predictions are more difficult for \nMgE\ (see detailed discussion in Sections \ref{sec:fescLyC} and \ref{sec:LyC}).








\begin{figure*}
\center
	\includegraphics[angle=0,trim={0.0cm 0.8cm 0.1cm 0.0cm},clip=true,width=0.5\linewidth,keepaspectratio]{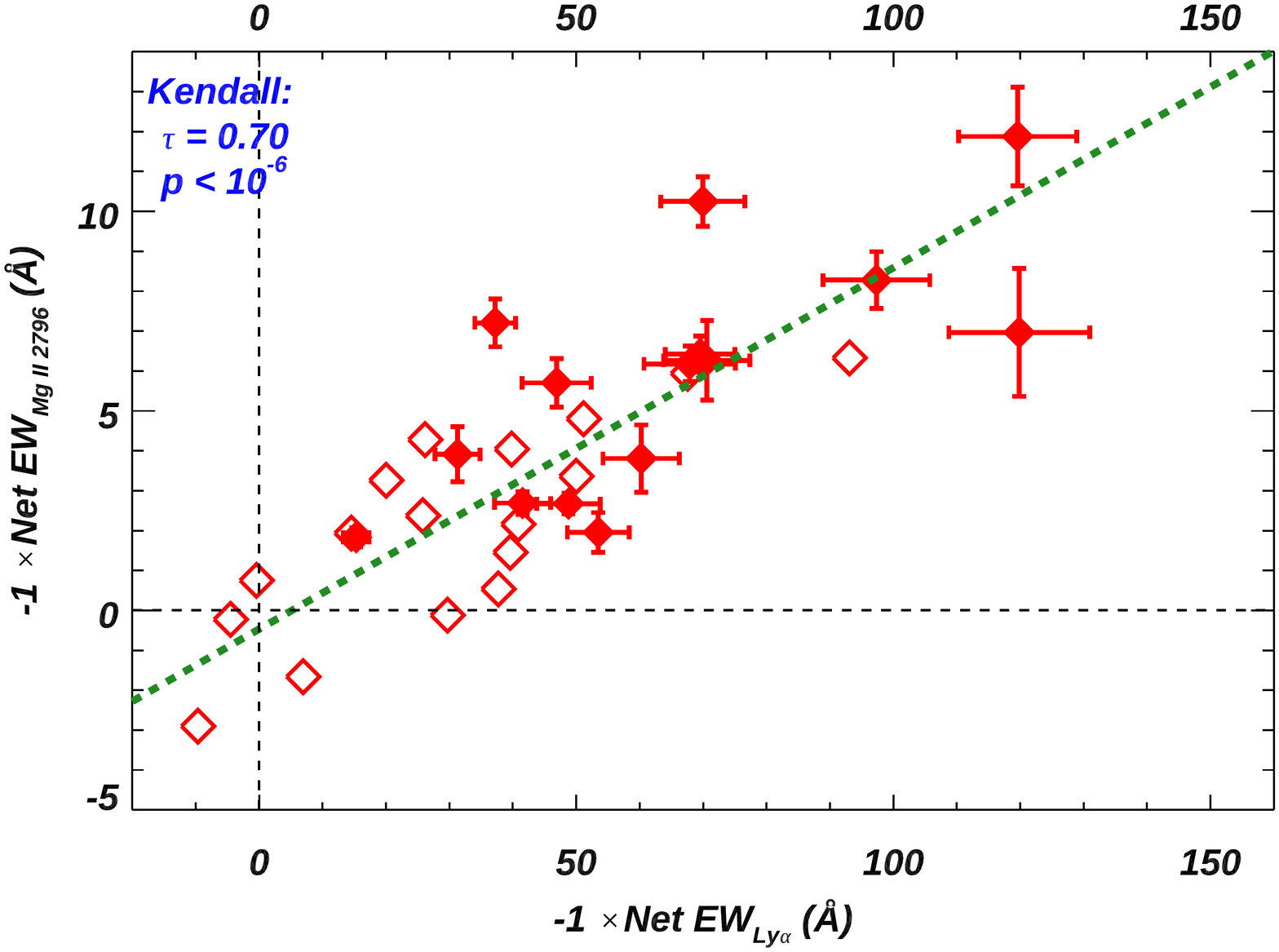}
	\includegraphics[angle=0,trim={0.0cm 0.8cm 0.1cm 0.0cm},clip=true,width=0.5\linewidth,keepaspectratio]{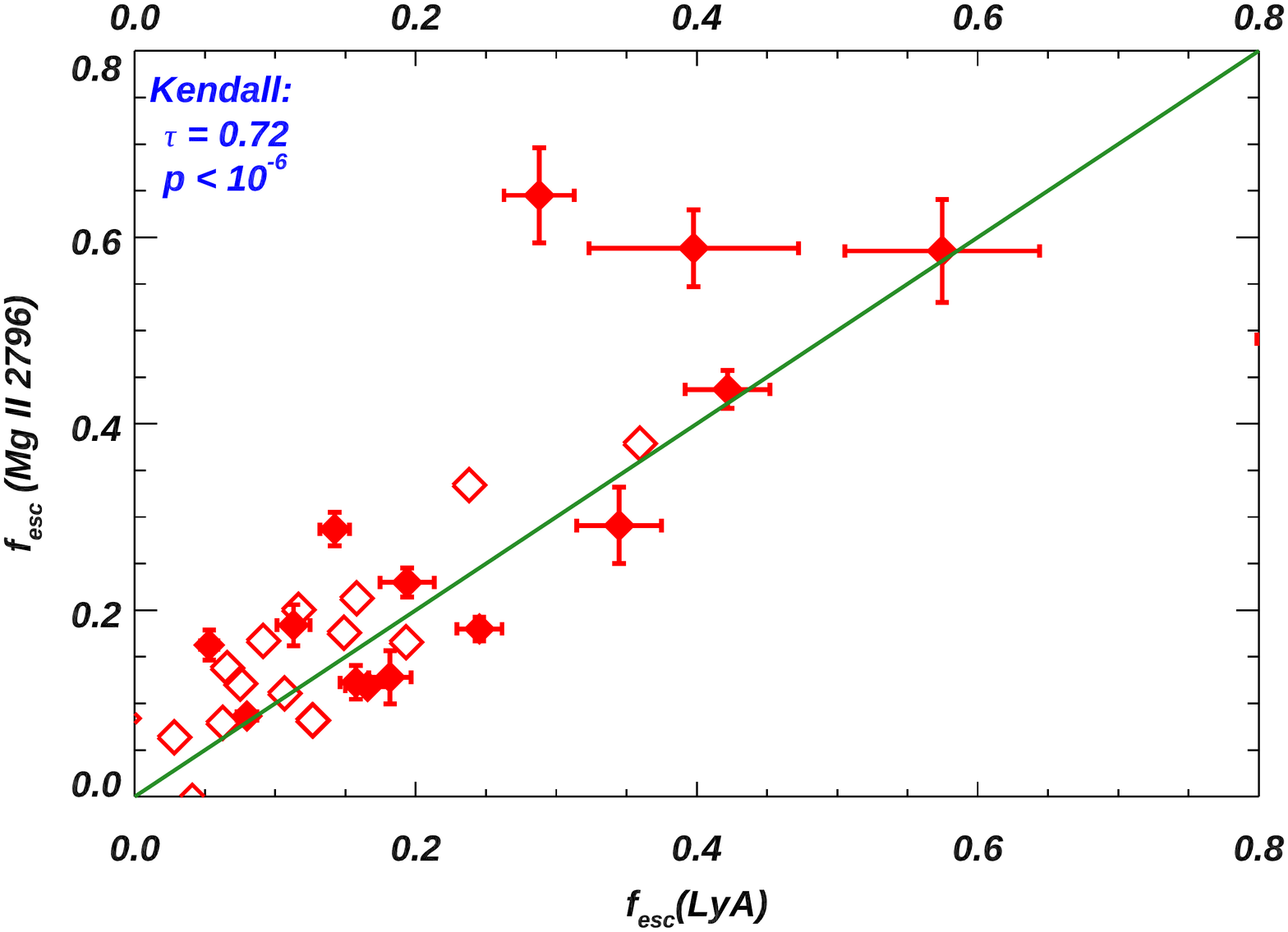}
	
\caption{\normalfont{Correlations between \mgii\ and \lya\ properties. Galaxies that are labelled as MgE and non-MgE from our sample are shown in filled and open symbols, respectively (Section \ref{sec:geometry}).  \textbf{Left:} The net EW from \mgii\ \ly 2796 and \lya\ are positively correlated. Each galaxy is shown as a dot with the cross representing its error bars. Galaxies with strong \mgii\ emission lines are at the top-right of the figure. The green dashed line represents the best linear fit. \textbf{Right:} The escape fraction of \mgii\ \ly 2796 and \lya\ are tightly correlated. The correlation coefficients from Kendall $\tau$ test are shown at the top-left corner in each panel. The green solid line represents the 1:1 correlation. See discussion in Section \ref{Sec:MgII}.} }
\label{fig:MgII-LyA1}
\end{figure*}

\subsection{The Method to Predict the Escape Fraction of LyC}
\label{sec:fescLyC}
As discussed in numerous previous publications \citep[e.g.,][]{Zackrisson13, Reddy16b,  Chisholm20, Kakiichi21,Saldana-Lopez22}, the escape of LyC photons can be described as a partial-covering geometry: 

\begin{equation}\label{eq:clumpy}
\begin{aligned} 
    f_\text{esc}(\text{LyC}) &=  C_{f}(\text{H I})e^{-\tau_\text{thick}}\times 10^{-0.4A_\text{thick}} \\
        &+ [1-C_{f}(\text{H I})]e^{-\tau_\text{thin}}\times 10^{-0.4A_\text{thin}}
\end{aligned} 
\end{equation}
where \CFHI\ is the covering fractions for optically thick paths of \hi\ which is dominated by neutral gas, and \Athick\ and \Athin\ are the attenuation parameters for LyC photons at optically thick and optically thin paths, respectively. 

For galaxies in our LzLCS sample, \cite{Saldana-Lopez22} have found that the covering fraction of lower ionization lines (LIS, including \oi, \cii, \Siii) trace that of \hi. Given similar ionization potentials of \mgii\ to these lines, we adopt their best-fit linear correlation to estimate \CFHI\ as:

\begin{equation}\label{eq:CF}
\begin{aligned} 
   C_{f}(\text{H I}) = (0.63 \pm 0.19) C_{f}(\text{Mg II}) + (0.54 \pm 0.09)
\end{aligned} 
\end{equation}


For optically thick paths, we assume no LyC photons can escape (i.e., \Tthick\ and/or \Athick\ $\gg$ 1). Therefore, the first term in Equation (\ref{eq:clumpy}) is negligible. \Athin\ is related to the dust extinction at the LyC, for which we adopt the stellar extinction derived from SED fittings in \cite{Saldana-Lopez22}. They used Starburst99 template \citep{Leitherer99} and have assumed the extinction law from \cite{Reddy15, Reddy16b}. Therefore, we can rewrite Equation (\ref{eq:clumpy}) as:


\begin{equation}\label{eq:LyC}
\begin{aligned} 
    f^\text{LyC}_\text{esc,pd} = [1-C_{f}(\text{H I})]e^{-N(\text{H I})\sigma_{ph}} \times 10^{-0.4E(B-V)k(912)}
\end{aligned} 
\end{equation}
where $f^\text{LyC}_\text{esc,pd}$ is the predicted absolute escape fraction of LyC, N(\hi) is the column density of neutral hydrogen, $\sigma_{ph}$ is the photoionization cross section of \hi\ at 912\angstrom, $E(B-V)$ is the stellar dust extinction from \cite{Saldana-Lopez22}, and k(912) is the total attenuation curve at the Lyman limit. Given the Reddy extinction law adopted in \cite{Saldana-Lopez22}, we have k(912) = 12.87. For other extinction laws, e.g., \cite{Cardelli89} and \cite{Calzetti00}, k(912) = 21.32 and 16.62, respectively.



As shown in \cite{Chisholm20} and \cite{Xu22b}, by assuming that \mgii\ and LyC photons escape from similar optically thin paths, the column density of \mgii\ [i.e., N(\mgii)] can be used to trace N(\hi) in a large range from DB to nearly IB regions:

\begin{equation}\label{eq:NHI}
\begin{aligned} 
    N(\text{H I})   &=  \alpha \times N(\text{Mg II}) 
\end{aligned} 
\end{equation}
where N(\text{Mg II}) can be calculated from the optical depth of \mgii\ as inferred from $R$ in Section \ref{sec:geometry}, and $\alpha$ = N(\mgii)/N(\hi) is the column density ratios predicted from CLOUDY models \citep{Xu22b}. $\alpha$ is dependent on the abundance ratio of [Mg/H] and the ionization, and has typical values $\sim$ 10$^{4}$ -- 10$^{5}$ for galaxies in our sample. Combining Equations (\ref{eq:clumpy}) to (\ref{eq:NHI}), we can calculate \fescLyCPred\ given the information of \mgii, metallicity, and dust.

In $\sim$ 20 galaxies with strong \mgii\ emission lines, this model predicted \fescLyCPred\ has been found to correlate well with the actual \fescLyC\ measured from the spectra \citep{Chisholm20, Xu22b}. Likewise, \cite{Katz22} also found that \fescLyCPred\ correlates with the actual \fescLyC\ for simulated high-redshift galaxies at different LOS (top-middle panel of their Figure 17). But their correlation contains a large scatter. We note that they adopt $\alpha$ as the abundance ratio of hydrogen to oxygen, i.e., $\alpha$ = 46$\frac{H}{O}$ \citep{Chisholm20}. This assumes the \mgii\ emission is found in neutral gas. This is not accurate since known LCEs commonly have high O$_{32}$ values and, thus, at least a fraction of the ISM (i.e., 1 - \CFHI) is density bounded. Therefore, \mgii\ in these galaxies should originate in regions where N(\hii) is non-negligible. Our adopted $\alpha$ from CLOUDY models in Equation \ref{eq:NHI} overcomes this problem \citep{Xu22b}. In Section \ref{sec:LyC}, we compare \fescLyCPred\ with the measured \fescLyC\ from \cite{Flury22a} based on the HST COS/G140L spectra and SED fittings.


\section{Results}
\label{sec:results}


\subsection{Estimates of the Escape Fraction of \mgii}
\label{Sec:MgII}


From Sections \ref{sec:geometry} to \ref{sec:fescMgII}, we show how to derive \fescMgII. The resulting values are listed in the last column of Table \ref{tab:MgII}.  Furthermore, in Figure \ref{fig:MgII-LyA1}, we present the correlations between \mgii\ and \lya. In the left panel, we compare the net EW (i.e., the summed EW from both emission and absorption features) between \mgii\ and \lya. To be consistent with the literature, we have multiplied the net EW by -1 to allow galaxies with strong emission lines to be at the top-right corner, while galaxies with strong absorption lines to be at the bottom-left corner. We find a strong positive correlation between the net EW of \mgii\ and \lya. This is as expected since both are resonant lines and should follow similar radiative transfer processes when travelling out of the galaxies. The best fit linear correlation is (show as the green dashed line):

\begin{equation}\label{eq:MgII-LyA}
\begin{aligned} 
    \text{net EW(Mg II)} & = a + b \times \text{net EW}(\text{Ly}\alpha)\\
    a & = -0.468_{-0.157}^{+0.157}\\
    b & = 0.091_{-0.003}^{+0.003}
\end{aligned}
\end{equation}

Similar but less significant trends between EW(\mgii) and EW(\lya) have also been published in \cite{Henry18} and \cite{Xu22b}, where they only focused on strong \mgii\ emitters. 


In the right panel of Figure \ref{fig:MgII-LyA1}, we compare the derived \fescMgII\ with the escape fraction of \lya\ (\fescLyA). The latter is derived from each galaxy's HST/COS spectra in \cite{Flury22a}. The correlation is significant ($p$ $<$ 10$^{-6}$) with scatter. We find most of the galaxies follow the 1:1 correlation shown as the solid green line, which suggests \fescLyA\ $\simeq$ \fescMgII. This is consistent with the results in \cite{Henry18} and \cite{Xu22b}, which also found that \fescMgII\ and \fescLyA\ values are of the same order. This supports the scenario where \mgii\ and \lya\  mainly escape from optically thin (or DB) holes in ISM likely in a single flight \citep[e.g.,][]{Gazagnes18, Chisholm20, Saldana-Lopez22}. Thus, the path lengths of \mgii\ and \lya\ photons travelling out of the galaxy are similar, and the resulting escape fractions are close for both lines. One possible scenario is that there is zero dust in the optically thin paths. Future spatially resolved observations can solve this puzzle. This includes our Lyman-alpha and Continuum Origins Survey (LaCOS, HST-GO 17069, PI: Hayes), which aims to spatially resolve the \lya\ emission, dust, and stellar population for 41 out of 66 LzLCS galaxies by HST imaging. 




For all figures in this section, we show Kendall $\tau$ coefficients and the probability of a spurious correlation (p values) at the top-left corner. In the Kendall test, we have accounted for the upper limits (if any) following \cite{Akritas96}. We have also tested the correlations between the scatters in each figure with other galaxy properties, including metallicity, internal dust extinction, SFR surface density, and stellar mass. However, we do not find significant correlations.




\subsection{Estimates of the Escape Fraction of LyC from \mgii}
\label{sec:LyC}

\begin{figure}
\center

	\includegraphics[angle=0,trim={0.0cm 0.8cm 0.0cm 0.0cm},clip=true,width=1.0\linewidth,keepaspectratio]{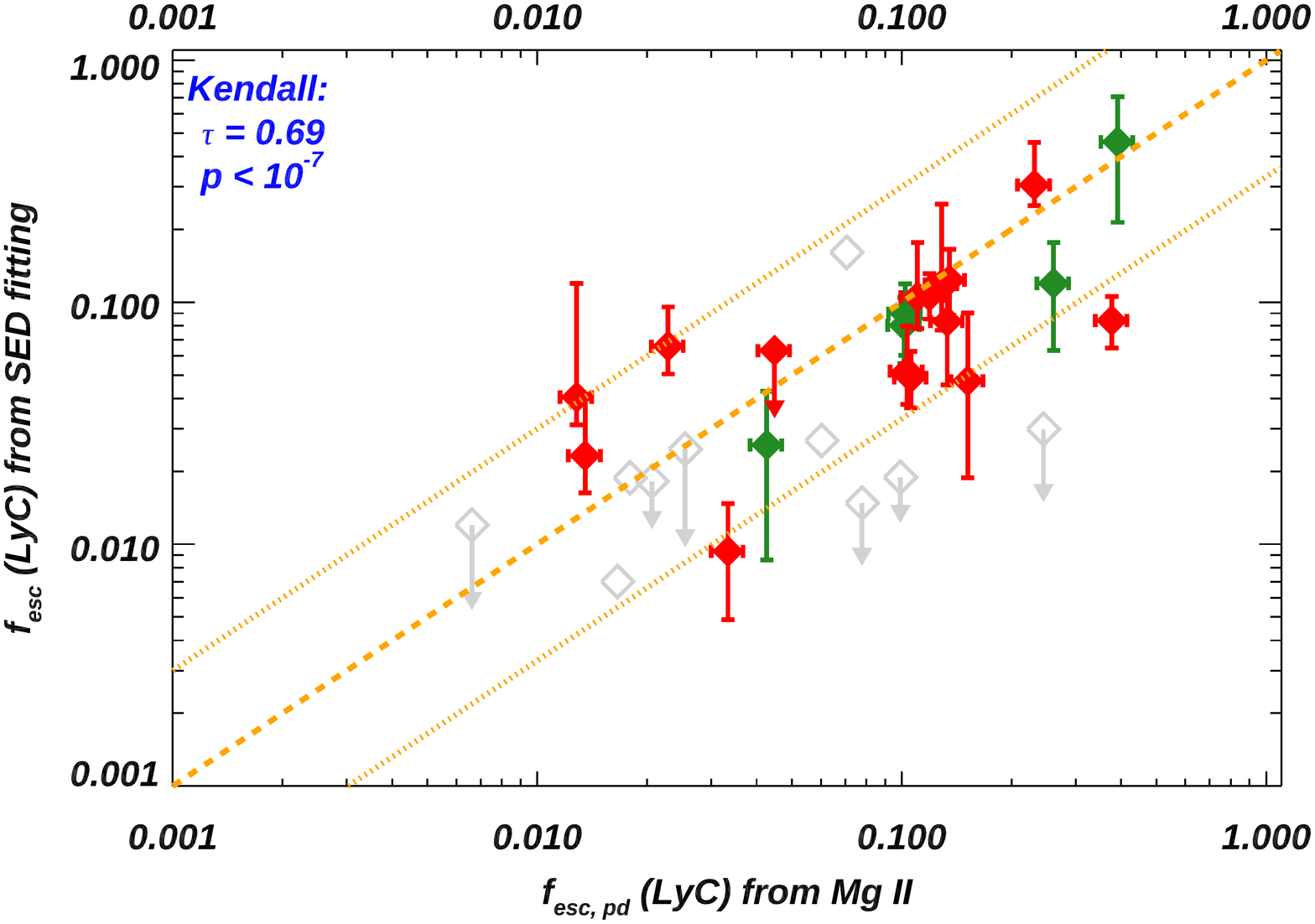}

\caption{\normalfont{Comparisons of measured \fescLyC\ with the predicted one from \mgii\ \ly 2796 emission lines. Galaxies that are labelled as MgE and non-MgE from our sample are shown in filled-red and open-gray symbols, respectively (Section \ref{sec:geometry}). Galaxies that are determined to be non-Lyman-continuum-emitter \citep{Flury22a} are shown as upper limits. The orange dotted lines are to show the factor of 3 scatter around the 1:1 relationship (orange dashed line). We also show 5 galaxies from \cite{Guseva20} as green colors. See discussion in Section \ref{sec:LyC}. } }
\label{fig:LyCFromMgII}
\end{figure}

In Figure \ref{fig:LyCFromMgII}, we compare \fescLyCPred\ derived in Section \ref{sec:fescLyC} with the \fescLyC\ values derived from the HST/COS spectra \citep[based on UV continuum fittings,][]{Flury22a}.  We draw galaxies classified as \MgE\ and \nMgE\ as filled and open symbols, respectively. We also include 5 galaxies from \cite{Guseva20}, which have high-quality VLT/X-Shooter observations as well as direct LyC measurements. We derive \fescLyCPred\ in the same way as in Section \ref{sec:fescLyC}, and remeasure their \fescLyC\ from HST/COS spectra using the same methodology in \cite{Flury22a}.

First of all, there is a strong correlation between the predicted \fescLyC\ from \mgii\ and measured ones, given the probability of a spurious correlation $p < 10^{-7}$. This highlights the power of using \mgii\ to trace LyC. Given the scatter around the 1:1 relationships line (orange dashed line), our predicted \fescLyC\ values are accurate within a factor of 3. Considering only the non-MgEs (gray-open symbols), the correlation is less significant. This can be because their \mgii\ emission lines are affected by absorption features, and the derived $\tau_{2803,thin}$ and \CF\ from Equation (\ref{eq:MgIItau}) is more uncertain. 

For some of the \nMgE, their \mgii\ spectra show significant resonant scattering or absorption signatures. These include galaxies showing double peaks in each \mgii\ emission lines (J1310+2148, J1244+0215, J0826+1820, and maybe in J1235+0635), and galaxies showing strong absorption in \mgii\ (J0723+4146, J0940+5932, J1346+1129, J1314+1048, J0957+2357). These galaxies should have optically thicker clouds in/around the galaxy, and our measurements of emission line flux of \mgii\ are also uncertain. Thus, no meaningful predictions through \mgii\ emission lines can be made, and we have excluded these galaxies from Figure \ref{fig:LyCFromMgII}. We note that, among these galaxies, only one (J1310+2148, \fescLyC\ $\sim$ 1.6\%) has small amount of LyC detected from HST/COS spectra, and others show non-detections of LyC (see Figures \ref{fig:MgII-1} and \ref{fig:MgII-2}). Thus, these non-MgEs are more similar to galaxies that are cosmologically irrelevant to EoR (\fescLyC\ $\ll$ 1\%). Therefore, precise estimates of their \fescLyC\ are less important for our understanding of SF galaxies contributing to the reionization. Furthermore, when selecting new LyC emitters for future observations, one can also exclude similar non-MgEs based on the absorption and/or scattering features in \mgii\ line profiles.




In the future, we plan to perform detailed radiative transfer models to account for the escape of \mgii\ out of the galaxy and link it to the escape of \lya\ and LyC (Carr et al. in preparation). We can then model and separate the emission and absorption features from the observed \mgii\ spectra. This will be particularly helpful to make more realistic predictions of \fescLyCPred\ for these galaxies labelled as \nMgE.

Overall,  our derived \fescLyCPred\ from \mgii, metallicity, and dust can correctly trace the measured \fescLyC\ within a factor of $\sim$ 3 for MgE. This is consistent with previous studies in \cite{Chisholm20, Xu22b}. We conclude that \mgii\ emission lines along with dust can be used to predict the escape of LyC photons in MgEs, but we need additional information to do so in non-MgEs (e.g., detailed radiative transfer models). 

\begin{table*}
	\centering
	\caption{Measurements from Optical Spectra for the Comparison Sample}
	\label{tab:MgII}
	\begin{tabular}{lcccccccccc} 
		\hline
		\hline
		Object 	    & O$_{32}$   & O/H     &	F$_{2796}^{Emi}$    & F$_{2803}^{Emi}$  & |EW$_{2796}^{Emi}$|  & |EW$_{2803}^{Emi}$|   & |EW$_{2796}^{Abs}$|  & |EW$_{2803}^{Abs}$| &  Label & \fescMgII\  \\
		\hline
		   (a)  &(b)  &(c)  &(d)  &(e)  &(f)  &(g)  &(h)  &(i)  &(j) &(k)   \\
		\hline

J1033+6353 & 3.4 & 8.2 & 50.2$\pm{7.0}$ & 34.9$\pm{7.0}$ & 4.7$\pm{0.8}$ & 3.6$\pm{0.8}$ & 0.9$\pm{0.3}$ & 0.0$\pm{0.0}$ & MgE & 0.29$\pm{0.04}$\\
J0917+3152 & 2.0 & 8.5 & 16.2$\pm{7.4}$ & 13.5$\pm{6.8}$ & 1.1$\pm{0.4}$ & 0.9$\pm{0.6}$ & 1.2$\pm{0.4}$ & 0.9$\pm{0.3}$ & non-MgE & 0.06$\pm{0.03}$\\
J1327+4218 & 3.3 & 8.2 & 38.2$\pm{4.4}$ & 22.7$\pm{4.8}$ & 6.0$\pm{0.6}$ & 4.1$\pm{1.2}$ & 0.3$\pm{0.08}$ & 0.5$\pm{0.1}$ & MgE & 0.18$\pm{0.02}$\\
J1410+4345 & 8.3 & 8.0 & 10.8$\pm{2.4}$ & 7.8$\pm{2.6}$ & 7.3$\pm{1.6}$ & 5.6$\pm{2.0}$ & 0.3$\pm{0.09}$ & 0.0$\pm{0.0}$ & MgE & 0.13$\pm{0.03}$\\
J1158+3125 & 1.8 & 8.4 & 75.5$\pm{5.2}$ & 38.6$\pm{4.8}$ & 3.2$\pm{0.2}$ & 1.7$\pm{0.2}$ & 0.6$\pm{0.2}$ & 0.3$\pm{0.09}$ & MgE & 0.18$\pm{0.01}$\\
J1235+0635 & 3.4 & 8.4 & 16.1$\pm{1.6}$ & 10.6$\pm{1.4}$ & 2.3$\pm{0.2}$ & 1.5$\pm{0.2}$ & 0.5$\pm{0.1}$ & 0.0$\pm{0.0}$ & MgE & 0.16$\pm{0.02}$\\
J1248+1234 & 3.4 & 8.2 & 116.3$\pm{5.0}$ & 62.3$\pm{4.6}$ & 9.6$\pm{0.6}$ & 5.0$\pm{0.4}$ & 1.3$\pm{0.4}$ & 0.5$\pm{0.1}$ & MgE & 0.44$\pm{0.02}$\\
J1517+3705 & 2.5 & 8.3 & 41.8$\pm{1.2}$ & 26.7$\pm{1.4}$ & 7.4$\pm{0.4}$ & 4.2$\pm{0.2}$ & 1.2$\pm{0.2}$ & 0.7$\pm{0.2}$ & MgE & 0.12$\pm{0.003}$\\
J0122+0520 & 5.7 & 7.8 & 27.7$\pm{2.6}$ & 17.5$\pm{2.4}$ & 6.4$\pm{1.0}$ & 4.8$\pm{1.2}$ & 0.1$\pm{0.03}$ & 0.3$\pm{0.09}$ & MgE & 0.59$\pm{0.06}$\\
J1301+5104 & 3.3 & 8.3 & 31.4$\pm{5.6}$ & 16.4$\pm{5.0}$ & 5.0$\pm{0.8}$ & 3.1$\pm{1.0}$ & 0.7$\pm{0.2}$ & 0.3$\pm{0.09}$ & non-MgE & 0.18$\pm{0.03}$\\
J1648+4957 & 3.3 & 8.2 & 35.4$\pm{0.4}$ & 23.9$\pm{0.4}$ & 21.3$\pm{0.2}$ & 16.1$\pm{0.2}$ & 3.5$\pm{0.2}$ & 0.0$\pm{0.0}$ & MgE & 0.49$\pm{0.006}$\\
J0911+1831 & 1.6 & 8.1 & 50.4$\pm{7.4}$ & 34.4$\pm{8.2}$ & 2.9$\pm{0.4}$ & 2.0$\pm{0.4}$ & 1.0$\pm{0.3}$ & 0.5$\pm{0.2}$ & MgE & 0.12$\pm{0.02}$\\
J0113+0002 & 2.3 & 8.3 & 42.9$\pm{3.0}$ & 23.1$\pm{2.6}$ & 5.0$\pm{0.6}$ & 2.9$\pm{0.4}$ & 1.1$\pm{0.3}$ & 0.5$\pm{0.1}$ & MgE & 0.59$\pm{0.04}$\\
J1133+4514 & 3.6 & 8.0 & 83.9$\pm{6.4}$ & 43.1$\pm{6.6}$ & 7.3$\pm{0.6}$ & 3.8$\pm{0.6}$ & 0.1$\pm{0.04}$ & 0.0$\pm{0.0}$ & MgE & 0.65$\pm{0.05}$\\
J0811+4141 & 8.1 & 7.9 & 45.7$\pm{4.2}$ & 28.0$\pm{4.0}$ & 12.9$\pm{1.2}$ & 7.4$\pm{1.0}$ & 1.0$\pm{0.3}$ & 0.0$\pm{0.0}$ & MgE & 1.00$\pm{0.23}$\\
J0958+2025 & 5.2 & 7.8 & 21.2$\pm{5.0}$ & 8.4$\pm{4.2}$ & 8.2$\pm{2.6}$ & 4.4$\pm{2.8}$ & 2.2$\pm{0.7}$ & 0.5$\pm{0.1}$ & non-MgE & 0.11$\pm{0.03}$\\
J1310+2148 & 1.6 & 8.4 & 18.7$\pm{2.0}$ & 12.4$\pm{1.6}$ & 2.1$\pm{0.4}$ & 1.4$\pm{0.2}$ & 1.6$\pm{0.5}$ & 0.3$\pm{0.09}$ & non-MgE & 0.06$\pm{0.007}$\\
J0047+0154 & 2.9 & 8.0 & 41.8$\pm{2.8}$ & 31.0$\pm{3.0}$ & 4.0$\pm{0.2}$ & 3.1$\pm{0.2}$ & 1.3$\pm{0.2}$ & 0.8$\pm{0.2}$ & MgE & 0.23$\pm{0.02}$\\
J1038+4527 & 1.5 & 8.4 & 59.9$\pm{8.4}$ & 35.4$\pm{9.0}$ & 2.8$\pm{0.4}$ & 1.7$\pm{0.4}$ & 0.8$\pm{0.2}$ & 0.6$\pm{0.2}$ & non-MgE & 0.14$\pm{0.02}$\\
J1246+4449 & 3.4 & 8.0 & 72.4$\pm{4.2}$ & 41.4$\pm{4.0}$ & 10.8$\pm{0.6}$ & 6.2$\pm{0.6}$ & 0.5$\pm{0.2}$ & 0.1$\pm{0.04}$ & MgE & 0.29$\pm{0.02}$\\
J0834+4805 & 3.6 & 8.2 & 55.6$\pm{1.4}$ & 44.7$\pm{1.4}$ & 7.4$\pm{0.4}$ & 6.1$\pm{0.2}$ & 1.0$\pm{0.2}$ & 0.0$\pm{0.0}$ & MgE & 0.09$\pm{0.002}$\\
J1244+0215 & 3.6 & 8.2 & 64.3$\pm{7.6}$ & 40.4$\pm{7.4}$ & 2.8$\pm{0.6}$ & 1.8$\pm{0.6}$ & 0.7$\pm{0.2}$ & 0.5$\pm{0.2}$ & non-MgE & 0.08$\pm{0.009}$\\
J1130+4935 & 3.4 & 8.3 & 15.2$\pm{2.0}$ & 7.4$\pm{1.8}$ & 6.3$\pm{0.8}$ & 3.1$\pm{0.8}$ & 1.5$\pm{0.4}$ & 0.9$\pm{0.3}$ & non-MgE & 0.38$\pm{0.05}$\\
J0129+1459 & 1.6 & 8.4 & 33.7$\pm{7.6}$ & 15.8$\pm{8.2}$ & 3.3$\pm{0.8}$ & 1.7$\pm{1.2}$ & 1.8$\pm{0.5}$ & 1.2$\pm{0.4}$ & non-MgE & 0.17$\pm{0.04}$\\
J0036+0033 & 10.5 & 7.8 & 14.4$\pm{2.2}$ & 7.5$\pm{2.8}$ & 7.7$\pm{1.8}$ & 5.3$\pm{2.4}$ & 1.3$\pm{0.4}$ & 1.0$\pm{0.3}$ & non-MgE & 0.20$\pm{0.03}$\\
J0926+3957 & 2.2 & 8.2 & 12.3$\pm{2.2}$ & 6.1$\pm{1.8}$ & 4.1$\pm{0.8}$ & 1.8$\pm{0.6}$ & 0.8$\pm{0.3}$ & 1.0$\pm{0.3}$ & non-MgE & 0.17$\pm{0.03}$\\
J0826+1820 & 4.0 & 8.3 & 7.5$\pm{2.6}$ & 3.4$\pm{2.2}$ & 2.8$\pm{1.0}$ & 1.8$\pm{1.4}$ & 0.4$\pm{0.1}$ & 1.2$\pm{0.4}$ & non-MgE & 0.12$\pm{0.04}$\\
J0912+5050 & 3.0 & 8.2 & 25.3$\pm{3.2}$ & 19.6$\pm{3.6}$ & 4.4$\pm{0.6}$ & 3.5$\pm{0.6}$ & 0.4$\pm{0.1}$ & 0.1$\pm{0.04}$ & non-MgE & 0.21$\pm{0.03}$\\
J0814+2114 & 1.2 & 8.1 & 39.4$\pm{9.2}$ & 28.7$\pm{10.0}$ & 1.4$\pm{0.4}$ & 0.9$\pm{0.4}$ & 0.7$\pm{0.2}$ & 0.3$\pm{0.08}$ & non-MgE & 0.06$\pm{0.01}$\\
J0723+4146 & 3.2 & 8.2 & 27.6$\pm{6.4}$ & 16.9$\pm{6.6}$ & 4.9$\pm{1.2}$ & 3.6$\pm{1.6}$ & 1.6$\pm{0.5}$ & 0.9$\pm{0.3}$ & non-MgE & 0.33$\pm{0.08}$\\
J0940+5932 & 1.5 & 8.4 & \dots & \dots & \dots & \dots & 5.9$\pm{0.2}$ & 4.2$\pm{0.4}$ & non-MgE & \dots\\
J1346+1129 & 1.1 & 8.3 & 72.1$\pm{11.0}$ & 68.7$\pm{12.8}$ & 2.3$\pm{0.4}$ & 2.2$\pm{0.6}$ & 2.6$\pm{0.4}$ & 2.3$\pm{0.4}$ & non-MgE & 0.11$\pm{0.02}$\\
J1314+1048 & 1.1 & 8.3 & 25.0$\pm{5.6}$ & 27.4$\pm{8.0}$ & 1.4$\pm{0.2}$ & 1.5$\pm{0.6}$ & 3.0$\pm{0.4}$ & 2.2$\pm{0.4}$ & non-MgE & 0.05$\pm{0.01}$\\
J0957+2357 & 0.5 & 8.4 & \dots & \dots & \dots & \dots & 2.9$\pm{0.4}$ & 2.6$\pm{0.4}$ & non-MgE & \dots\\

		\hline
		\hline
	\multicolumn{11}{l}{%
  	\begin{minipage}{16.5cm}%
	\textbf{Note.} --Measurements from the optical spectra for galaxies in our sample. Galaxies are ordered by decreasing \fescLyC\ derived in \cite{Flury22a} (the same order as Figures \ref{fig:MgII-1} and \ref{fig:MgII-2}). The columns are: (b) Flux ratio between [\oiii] \ly5007 and [\oii] \ly3727; (c) Gas phase metallicity in the form of 12+log(O/H); (d) and (e) Measured emission line flux of \mgii\ \ly\ly 2796, 2803 lines in units of 10$^{-17}$ ergs s$^{-1}$ cm$^{-2}$, respectively; (f) and (g): Measured rest-frame EW in units of \angstrom\ for the emission part from the \mgii\ doublet (see Section \ref{sec:Basic}); (h) and (i) Measured rest-frame EW in units of \angstrom\ for the absorption part from the \mgii\ doublet; (j) Labels based on the \mgii\ line profiles, i.e., MgE = \mgii\ emitter, non-MgE = \mgii\ non-emitter (see Section \ref{sec:geometry}); and (k): the derived escape fraction for \mgii\ \ly2796 (see Section \ref{sec:fescMgII}).

  	\end{minipage}%
	}\\
	\end{tabular}
	\\ [0mm]
	
\end{table*}


















\section{Conclusion and Future Work}
\label{sec:Conclusion}

We present the analyses of \mgii\ spectra for 34 galaxies chosen from the LzLCS sample. These galaxies have published HST/COS data for their LyC and \lya\ spectral regions, and we have obtained higher S/N and resolution spectra (than SDSS) for their \mgii\ regions. 

While previous studies of \mgii\ in Lyman Continuum Emitter (LCE) candidates have only focused on \mgii\ emitters (\MgE), galaxies in our sample have \mgii\ profiles ranging from strong emission to P-Cygni profiles, then to pure absorption. We find there is a significant trend ($p$ = 0.0216) that galaxies detected as strong LCEs show larger EW(\mgii) in emission lines, while non-LCEs present larger EW(\mgii) in absorption.

We discuss the picket-fence geometry for the escape of \mgii\ photons from galaxies. While this geometry has been found to apply well to galaxies categorized as MgE, it has limitations in the case of non-MgE. We then discuss how to use the CLOUDY photoionization models to help derive the escape fraction of \mgii\ (\fescMgII) from the optical spectra. For all galaxies in our sample, we find \fescMgII\ correlates with the escape fraction of \lya. We also show that the net equivalent width of \mgii\ and \lya\ are tightly correlated for both MgEs and non-MgEs. 

We also discuss the methods to predict the escape fraction of LyC (\fescLyCPred) from the measurements of \mgii, metallicity, and dust. We show that the predicted \fescLyCPred\ correlates well with the actual \fescLyC\ derived from the HST/COS spectra within a factor of $\sim$ 3. For non-MgEs, the correlation is less significant. This is because the absorption features in \mgii\ spectra for non-MgE complicate our measurements of \mgii\ emission lines. Additional information, e.g., from radiative transfer models, may help solve this problem.

In the future, one can apply the \mgii\ correlations to various different studies, including: 1). We will perform detailed radiative transfer models to account for the escape of \mgii\ from the galaxy (Carr et al. in preparation). This will be especially helpful for the cases of non-MgE, where the clouds in/around the galaxy are not optically thin to \mgii. 2) For high-z galaxies, one can adopt the observed \mgii\ features to estimate the intrinsic amount of \lya, which can be severely attenuated by the neutral IGM \citep[e.g.,][]{Mason18}. Thus, with the aid of \mgii, one can get more accurate estimates of the IGM neutral fractions from \lya. 3). One can conduct similar analyses of \mgii\ in higher-redshift LCE candidates, whose \mgii\ emission lines are shifted into the observable bands of the James Webb Space Telescope (JWST). The \lya--\mgii\ correlations can be adopted to select \lya\ emitters that have detectable \mgii\ spectra, and the \mgii--LyC correlation can be used to predict \fescLyC\ in the case when LyC cannot be directly detected.

\clearpage
\begin{acknowledgements}
X.X. and A.H. acknowledge support from NASA STScI grants GO 15865. Observations reported here were obtained at the MMT Observatory, a joint facility of the University of Arizona and the Smithsonian Institution.

Support for this work was provided by NASA through grant number HST-GO-15626 from the Space Telescope Science Institute. This research is based on observations made with the NASA/ESA Hubble Space Telescope obtained from the Space Telescope Science Institute, which is operated by the Association of Universities for Research in Astronomy, Inc., under NASA contract NAS 5–26555. These observations are associated with program(s)  13744, 14635, 15341, 15626, 15639, and 15941. STScI is operated by the Association of Universities for Research in Astronomy, Inc. under NASA contract NAS 5-26555. 

Based on observations collected at the European Organisation for Astronomical Research in the Southern Hemisphere under ESO programme 106.215K.001.

The Low-Resolution Spectrograph 2 (LRS2) was developed and funded by the University of Texas at Austin McDonald Observatory and the Department of Astronomy and by Pennsylvania State University. We thank the Leibniz-Institut für Astrophysik Potsdam (AIP) and the Institut für Astrophysik Göttingen (IAG) for their contributions to the construction of the integral field units.

ASL acknowledge support from Swiss National Science Foundation. HA is supported by CNES.

\end{acknowledgements}
\facilities{HST (COS), MMT (Blue channel), APO (SDSS), VLT (X-Shooter), HET (LRS2)}
\software{
CLOUDY \citep[v17.01,][]{Ferland17}}


\typeout{} 
\bibliography{main}{}
\bibliographystyle{aasjournal}

\clearpage

\end{document}